\documentclass[aps,twocolumn,groupedaddress,amssymb]{revtex4}
\usepackage{latexsym}
\usepackage{amsbsy}
\usepackage{amsmath}
\usepackage{graphicx}
\usepackage{xcolor}
\usepackage{bm}
\usepackage{hyperref} 
\usepackage[capitalise]{cleveref} 
\hypersetup{
    colorlinks=true,
    linkcolor=black,
    urlcolor=blue,
    citecolor=blue
}
\usepackage{nameref}  

\begin{document}

\author{Indraneel Sinha\textsuperscript{1}}
\altaffiliation{These authors contributed equally to this work.}
\author{Saurav Sachin\textsuperscript{1}}
\altaffiliation{These authors contributed equally to this work.}
\author{Prashant Kumar\textsuperscript{2}}
\author{Atul Pandey\textsuperscript{3}}
\author{Bijoy Kumar Kuanr\textsuperscript{2}}
\author{Sujit Manna\textsuperscript{1}}
\altaffiliation{Contact author:smanna@physics.iitd.ac.in} 

\affiliation{\textsuperscript{1}Department of Physics, Indian Institute of Technology Delhi, Hauz Khas, New Delhi 110016, India}
\affiliation{\textsuperscript{2}Special Centre for Nanoscience, Jawaharlal Nehru University, New Delhi-110067, India}
\affiliation{\textsuperscript{3}Max Planck Institute of Microstructure Physics, Weinberg 2, 06120 Halle, Germany}

\preprint{APS/123-QED}

\title{Controlled spin-to-charge conversion in noncollinear antiferromagnet-based Py/Mn$_{3}$Pt heterostructures
}

\date{\today}

\begin{abstract}
Noncollinear antiferromagnets (NCAFs) have recently emerged as promising candidates for future spintronic technologies, offering ultrafast switching, negligible stray fields allowing dense packing, and robustness against external magnetic perturbations. When interfaced with ferromagnets (FMs), they can strongly influence interfacial exchange and spin–torque mechanisms that enable manipulating magnetic order and realizing functionalities beyond conventional heavy metals (HMs) based FM/HM heterostructures. 
Here, we perform a broadband ferromagnetic resonance (FMR) study to systematically investigate the magnetization dynamics and spin-to-charge conversion in permalloy (Py) and Mn$_3$Pt bilayers.
High-quality Py films provide a well-defined FMR spectra with a low Gilbert damping parameter ( $\alpha_{\mathrm{eff}} \approx 9.8 \times 10^{-3}$). We observe a pronounced enhancement of damping with intrinsic value $\alpha_{\mathrm{int}} \approx 3.1 \times 10^{-2}$ in the Py/Mn$_3$Pt bilayer, indicating efficient spin pumping into the NCAF layer. Frequency dependent linewidth analysis shows a predominantly Gilbert type damping in the bilayers and the corresponding effective spin-mixing conductance ( $g^{\uparrow\downarrow}_{\mathrm{eff}} \approx 4.8 \times 10^{18}$m$^{-2}$) is comparable to that of other high-performance antiferromagnetic heterostructures. These results are significant for establishing NCAFs as a candidate material for spin generation and highlights the potential of Py/Mn$_3$Pt bilayers for efficient and ultrafast spintronic applications.    

\end{abstract}
\maketitle


\section{Introduction}
Antiferromagnetic materials (AFs) have emerged as promising candidates for next-generation spintronics devices due to their ultrafast spin dynamics, robustness against external magnetic fields, and potential for low-power operations \cite{Han2018, Mangin2006, Chowdhury2023, bai2022antiferromagnetism}. Their integration enables novel spin transport phenomena such as the spin Hall effect, domain wall motion, and spin wave propagation, offering enhanced scalability and energy efficiency for future spin based technologies \cite{shiino2016antiferromagnetic,wu2024current}. In recent experimental advancements, noncollinear antiferromagnets (NCAFs) with a triangular spin structure have revealed robust optical and magnetotransport responses such as the anomalous Hall effect (AHE) \cite{NCAFM_AHE_Nat_Nakatsuji2015}, anomalous Nernst effect \cite{NCAFM_ANE_Nat_phy_Ikhlas2017,pandey_ane}, Kerr effect \cite{NCAFM_MOKE_nat_photo_Higo2018}, and magnetic spin Hall effect (MSHE) \cite{yang2017topological,busch2021spin,wu2020magneto}. This unexpected charge and spin transports in NCAFs are driven by the magnetic symmetries, leading to a nonzero Berry curvature in the momentum space \cite{Nayak2016,zuniga2023observation,li2023field}.

The family of manganese based NCAFs Mn$_{3}$X (X= Ga, Ge, Sn, Pt) have gained further prominence as they combine the properties of conventional ferromagnets (large optical and magneto-transport responses) with the advantages of AFs (non-volatility, negligible stray fields and ultrafast dynamics) \cite{Nagaosa2010, Nayak2016, Fukami2016,yang2017topological,poelchen2023long,pal2022setting}. These compounds are highly regarded as spin Hall materials, as the transverse spin currents possess unconventional components rather than the orthogonal relations observed in the conventional spin Hall effect. The strength of spin polarization and spin Hall angle is strongly influenced by the crystallographic orientations \cite{cao2023anomalous}, as demonstrated experimentally in Mn$_3$Pt \cite{Bai2021,cao2023anomalous} and Mn$_3$Ir/Ni$_{80}$Fe$_{20}$ films \cite{Holanda2020,zhang2016giant}. 
The Berry-curvature driven electrical and magneto-optical responses, which can be used as a read-out signal, are suppressed in the cubic members of the Mn$_3$X family (X = Pt, Ir) owing to their large magnetocrystalline anisotropy and the presence of degenerate non-topological spin configurations \cite{panda2025efficient,xie2022magnetization,yoon2025electrical,rimmler2025non}. 
Mn$_3$Pt, for instance, exhibits two distinct spin configurations, 
a non-topological T$_{2}$ spin structure, characterized by a helical spin configuration \cite{Bai2021,gurung2019anomalous} and a  topological phase T$_{1}$, in which the Mn spins lie in the (111) plane and point toward the centers of the surrounding triangles  (Fig. 1(a)) \cite{kota2008ab}. Although both the spin states are energetically favorable during growth, one can tune the growth parameters to obtain the non-trivial T\textsubscript{1} state, as discussed in our previous work \cite{sinha2025occurrence}. While the previous studies focused on the trivial T\textsubscript{2} state \cite{Bai2021,gurung2019anomalous}, the T$_{1}$ state is more interesting, as it can exhibit spontaneous anomalous Hall effect  persisting beyond 300 K. 


In this work, we investigate the magnetization dynamics and spin-to-charge conversion efficiency in billayers consisting of permalloy (Py) and Mn$_3$Pt with the triangular T$_{1}$ AF spin order. Magnetic and transport measurements confirm the antiferromagnetic ground state with pronounced AHE response directly originating from Berry-curvature-driven effects. The spin-sinking in Mn$_3$Pt is probed through thickness-dependent FMR studies and its effect on the magnetic parameters of the Py layer. The efficiency of Mn$_3$Pt as a spin sink also depends on its epitaxial integration with Py. To gain insight into this aspect, scanning tunneling microscopy (STM) is employed to investigate the surface morphology of different terminations. In addition, we report the STM investigation of the thickness-dependent evolution of the Py surface for the first time. Our findings aim to establish Mn$_3$Pt based bilayers as viable candidates for efficient spin-to-charge conversion, paving the way for next-generation antiferromagnetic spintronic devices.

\section{Experimental Methods}
Mn$_3$Pt and Py(Ni$_{80}$Fe$_{20}$)/Mn$_3$Pt thin films were grown in an ultrahigh vacuum sputtering chamber (Hind High Vacuum, India) with a base pressure of 2 × 10$^{-8}$ mbar. Mn$_3$Pt (20 nm) films were  deposited on Si via co-sputtering of Mn (99.9\%) and Pt (99.999\%) targets powered by direct current (DC) and radio frequency (RF) sources, respectively. During growth, the working pressure is maintained at 7.2 x 10$^{-3}$ mbar. The sputtering powers for Pt and Mn targets were set at 25 W and 60 W respectively. 
 In order to remove surface contaminants, silicon substrates were ultrasonically cleaned in sequence with acetone and isopropyl alcohol. The substrates were further degassed by a series of cyclic annealing up to 700°C. The growth temperature was maintained at 600°C, and the samples were subsequently post-annealed for one hour to improve their crystallinity. Ultrathin films of Py with thicknesses (t = 10, 11, 14, 16 nm) are grown on 20 nm thick Mn$_3$Pt films at 400 °C. An optimized growth rate of 0.4 \AA s$^{-1}$, confirmed by x-ray reflectivity (XRR) measurements, was employed to precisely control the thickness of the Py layers. Structural characterization was performed using X-ray diffraction (XRD) with a PANalytical X'Pert diffractometer with Cu-K$\alpha$ source ($\lambda = 1.5418 \, \text{\AA}$). 
 The surface morphology was analyzed using atomic force microscopy (AFM) (Oxford Instruments Asylum Research, MFP-3D system).
 All scanning tunneling microscopy (STM) measurements were carried out at room temperature using  Quazar STM equipped with a home-built active vibration cancellation system \cite{sinha2025magnetic,sinha2025anomalous}. The STM tips were prepared from polycrystalline PtIr and W wires, which were chemically etched and subsequently annealed at high temperature.
  The magnetization properties were measured using Quantum Design magnetic property measurement system (MPMS3).  
 Magnetotransport measurements are performed in a standard four-probe geometry using a Quantum Design PPMS. The magnetization dynamics were investigated using vector network analyser (VNA-Keysight, USA) based broadband FMR spectroscopy through a coplanar waveguide (CPW) placed between the poles of DC magnet transmission mode.  The CPWs effectively transmit microwave signals from a radio frequency (RF) source across a wide range of frequencies. A microwave RF field ($H_{RF}$) near the CPW excites the magnetic sample at resonance with suitable DC magnetic field and frequency. The sample was placed upside down on the signal line of the “S”-shaped CPW, ensuring that the RF magnetic field ($h_{rf}$) to be perpendicular to the external DC magnetic field. The fixed frequency FMR spectra  was acquired by sweeping the direct current (DC) magnetic field in the field sweep mode.

\begin{figure*}
    \centering
    \includegraphics[width=0.9\linewidth]{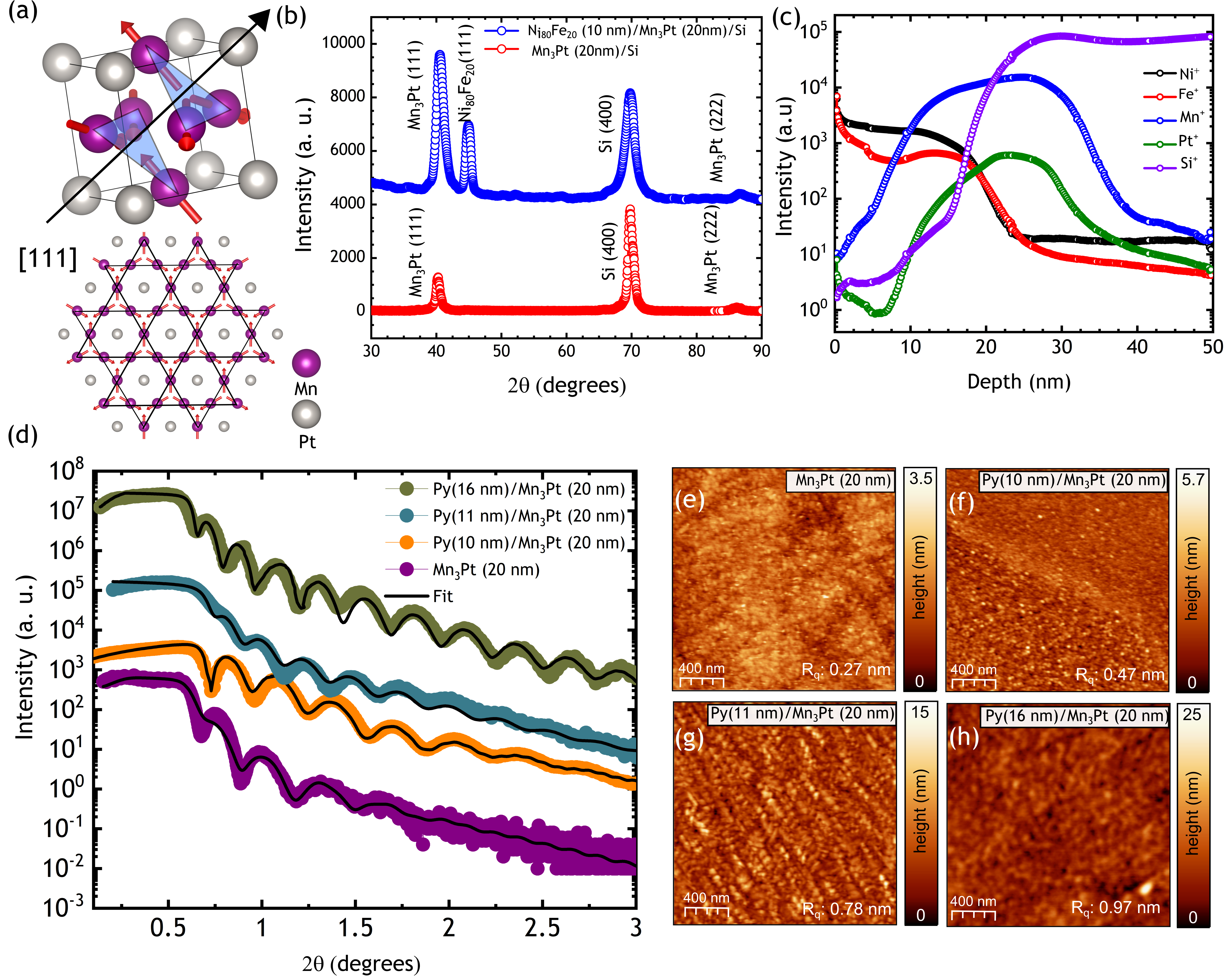}
    \caption{(a) Schematic illustration of the magnetic ground state of Mn$_{3}$Pt. (b) Out-of-plane X-ray diffraction (XRD) pattern measured on Mn$_{3}$Pt and Py/Mn$_{3}$Pt thin films. The strong diffraction peaks from the (111) and (222) planes demonstrate highly oriented growth along the (111) direction. (c) Positive TOF-SIMS depth profile of Py(16 nm)/Mn$_{3}$Pt(20 nm) (Intensity as a function of depth (nm)) confirms minimal inter-diffusion at interfaces. (d) Thickness dependent XRR spectra of Mn$_{3}$Pt and Py/Mn$_{3}$Pt heterostructures. Open symbols are experimental data points while solid lines represents fitting. (e-h) AFM surface topography of Mn$_{3}$Pt (20 nm) and Py (t)/Mn$_{3}$Pt (20 nm) bilayers on Si substrate over 2x2 $\mu$m$^{2}$ scanning area, where thickness of Py layer is varied from 10 nm to 16 nm.}
    \label{fig:characterization}
\end{figure*}

\section{Results and discussion}
\subsection{Growth and structural characterization}

X-ray diffraction (XRD) patterns of the Mn$_3$Pt and Py/Mn$_3$Pt films show a strong preferential orientation along the (111) crystallographic direction (Fig. 1(b))\cite{liu2018electrical}. 
The thicknesses of the Py(t)/Mn$_3$Pt bilayers, as accurately determined by XRR measurements 
indicate  a fixed Mn$_3$Pt layer thickness of  approximately 20 nm. 
To gain insight into the chemical depth distribution of the  bilayers, time-of-flight secondary ion mass spectrometry (TOF-SIMS) is used to investigate the depth profiling in a Py(16 nm)/Mn$_3$Pt(20 nm)/Si film, as shown in Fig. 1(c). The profile reveals a well-defined multilayer stack, where successive layers of Py, Mn$_3$Pt and Si substrate can be clearly identified from signals arising from Ni$^{+}$, Fe$^{+}$, Mn$^{+}$, Pt$^{+}$ and Si$^{+}$ respectively. The gradual transitions between these elemental profiles indicate well-defined interfaces and minimal inter-diffusion across the layers. The Mn and Pt signals are spatially confined within the intermediate region, consistent with the expected stoichiometry of Mn$_3$Pt. The presence of Ni and Fe signals diminishes before the onset of Mn and Pt, confirming the discrete layering. The Si$^{+}$ signal rises abruptly following the Mn$_3$Pt layer, further validating the successful fabrication of the heterostructure on the Si substrate. The interfacial roughness evaluated from XRR oscillations simulations (Fig. 1d), is found to be less than a nanometer. 
This shows high-quality contact at the interface, which is crucial for efficient spin pumping across the interface.
\begin{figure*}
\centering
\includegraphics[width=0.8\linewidth]{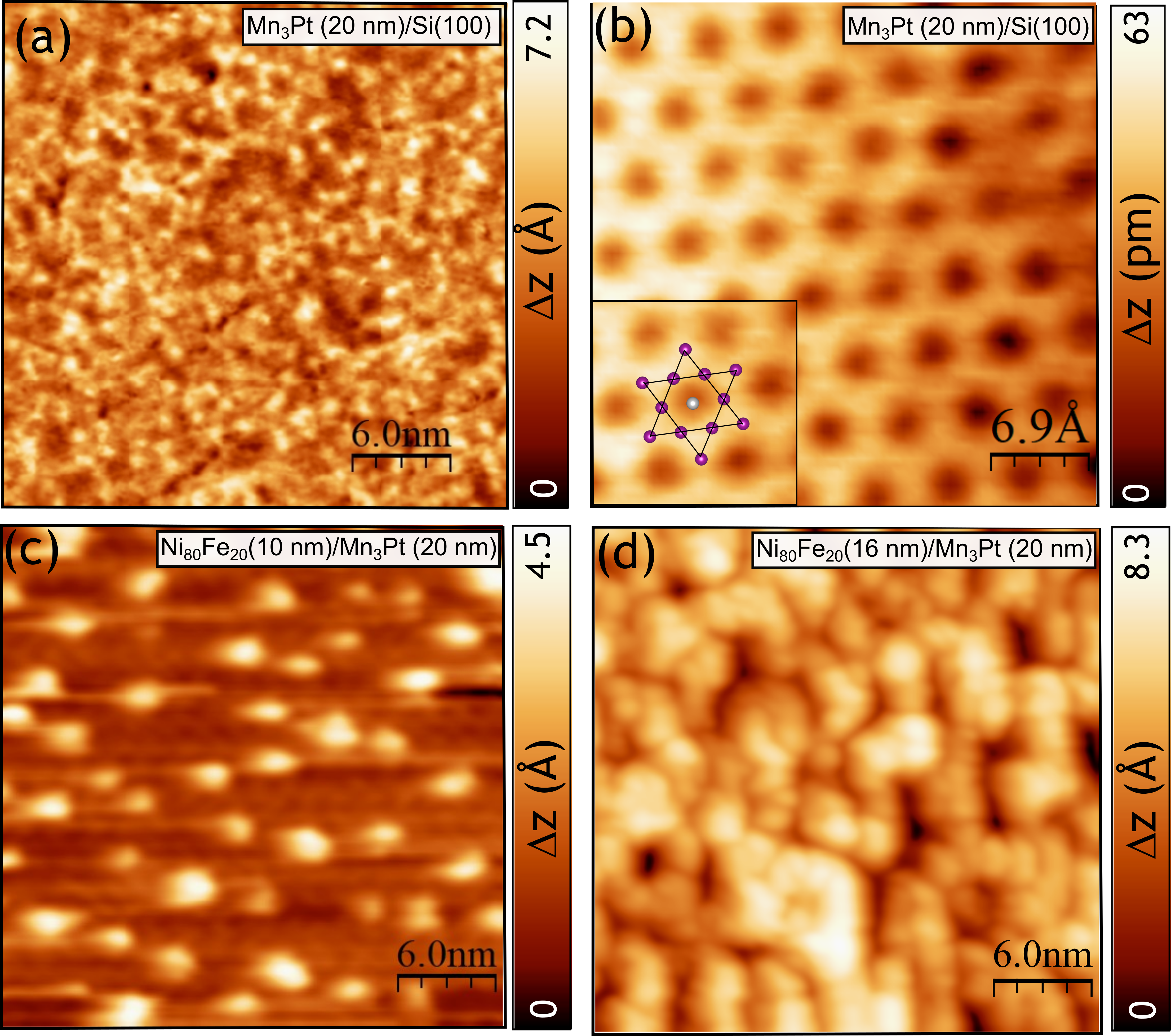}
\caption{(a) Constant-current STM topography image ($30 \times 30    \text{nm}^2$) showing the general surface morphology of 20 nm-thick Mn$_3$Pt(111) films epitaxially grown on Si(100). The image was acquired with a tunneling current of $I_s = 0.6~\text{nA}$ and sample bias of $V_{\text{bias}} = 500~\text{mV}$.
(b) High-resolution STM image ($3.5 \times 3.5  \text{nm}^2$) taken on a flat region, revealing the hexagonal arrangement surface atoms characterize by the (111)-terminated Mn$_3$Pt Kagome lattice. The inset shows a zoomed atomic structure overlaid with a Kagome lattice ball model. Imaging conditions: $I_s = 1~\text{nA}$, $V_{\text{bias}} = 135~\text{mV}$.
(c–d) Surface morphology of Py films with thicknesses of 10 nm and 16 nm, respectively, grown on top of 20 nm Mn$_3$Pt/Si(100). Images were acquired in constant-current mode with tunneling conditions of $I_s = 1.0~\text{nA}$ and $V_{\text{bias}} = 500$–$730~\text{mV}$.}
\label{fig:transport}
\end{figure*}

The surface morphology of Mn$_3$Pt and Py/Mn$_3$Pt films are further investigated using AFM. Topographic images presented in Fig. 1(e-h) reveal a smooth and homogeneous surface morphology for all samples, demonstrating topographical uniformity, which is critical to ensure reliable interface quality in heterostructures. Quantitative analysis of the root mean square (RMS) surface roughness shows values of 0.27 nm for Mn$_3$Pt (20 nm) and 0.47 nm, 0.78 nm \& 0.97 nm for Py(10 nm, 11 nm, 16 nm)/Mn$_3$Pt (20 nm) respectively. The low roughness of Py deposited on Mn$_3$Pt, consistent with XRR measurements, suggests a high degree of interfacial compatibility between layers, which is essential for spintronics applications. 

These measurements are complemented by the highly resolved STM topography ($30 \times 30$ nm\textsuperscript{2}) scans (Fig. 2). Mn$_3$Pt (111) surface shows an atomically flat termination, consistent with the large area AFM scans. Atomically resolved image acquired over $3.5 \times 3.5$ nm\textsuperscript{2} area shows the hexagonal arrangement of the Mn$_3$Pt (111) surface (Fig. 2b). The periodic atomic contrast modulation arises from variations in the local density of states at different atomic sites on the (111) termination. In the superimposed Kagome ball model shown in the inset, the bright atomic sites represent the Mn triangular sublattice, whereas the dark regions correspond to Pt atoms. The estimated distance between two dark sites is approximately 0.542 nm. Theoretically, the distance between two adjacent Pt atoms in the (111) plane is given by $\sqrt{2}a$, where $a$ is the lattice constant of Mn$_3$Pt. From this relation, the estimated in-plane lattice constant is found to be 0.384 nm. Such atomically flat termination is crucial for lattice-matched growth and efficient spin pumping.

High-resolution STM images ($30 \times 30$ nm\textsuperscript{2}) acquired on Py (10 nm)/Mn$_3$Pt (Fig.~2(c)) reveal the formation of clusters on a flat, well terminated surface. The top layers Py clusters are uniformly spatially distributed on the surface. The lateral dimension of the clusters vary between 1-3 nm. This granular morphology is attributed to the relatively low growth temperature of 400 $^\circ$C. Higher deposition temperature can stimulate the migration of grain boundaries and aids the coalescence of more and more grains which in turn increases surface roughness \cite{kumar2024influence}.  
However, the STM topography obtained on Py(16 nm)/Mn$_3$Pt ((Fig.~2(d)) shows coalescence of the grains, which leads to the formation of some large agglomerates. This has been previously observed in Py thin films grown on Si(100) substrate \cite{kharmouche2024structural}.

\begin{figure*}
\centering
\includegraphics[width=1.0\linewidth]{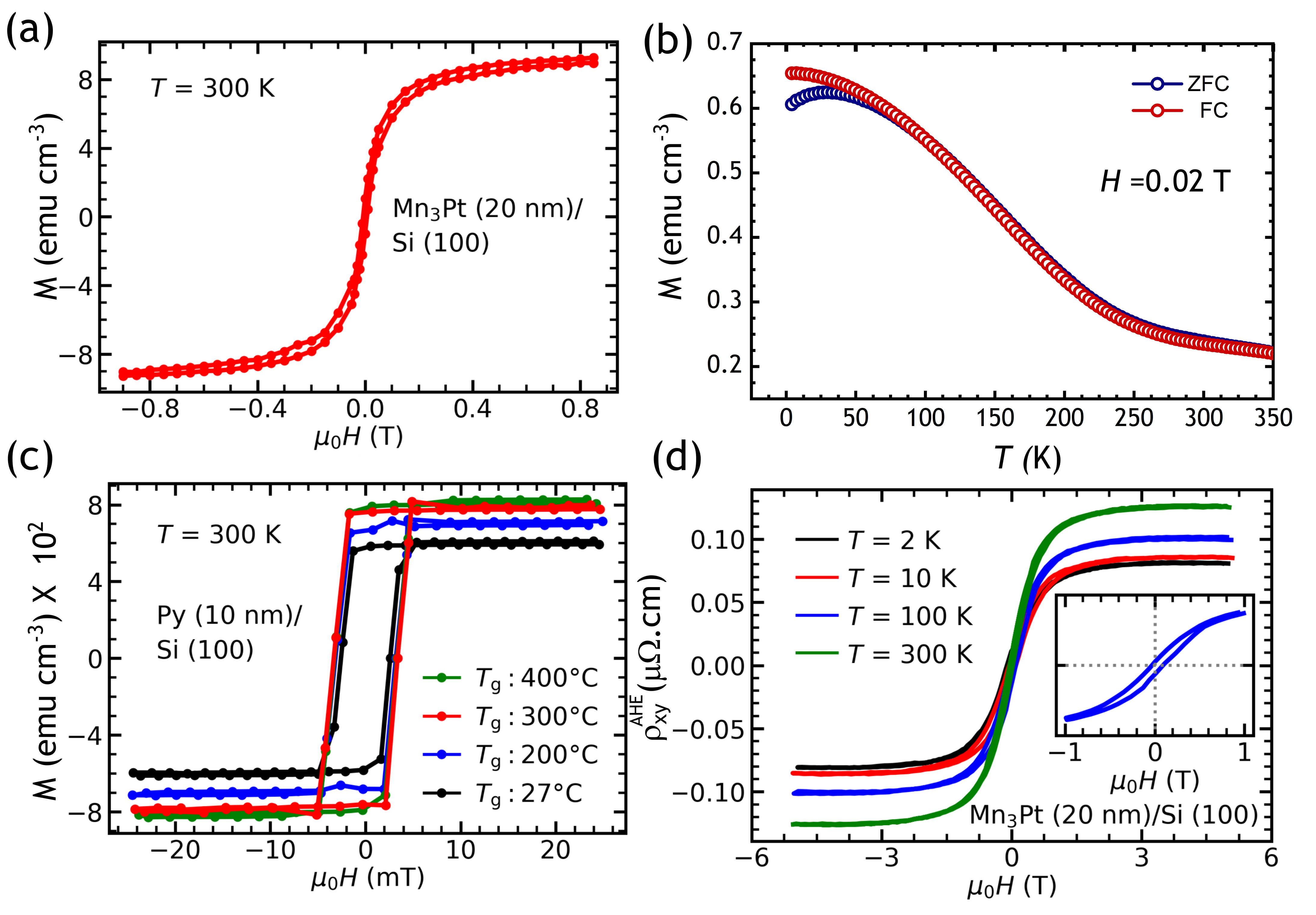}
\caption{Magnetization and magneto-transport studies: (a) 
Magnetization of Mn\textsubscript{3}Pt film (t = 20 nm) as a function of magnetic field measured at room temperature. The low saturation magnetization ($M_\mathrm{S} \approx 10~\text{emu}/\text{cm}^3$) originates from the uncompensated moments in the Kagome plane.
(b) Magnetization versus temperature measured in ZFC-FC mode under a perpendicular magnetic field of 200 Oe. (c) M-H loops of Py films grown at 27~$^\circ$C, 200~$^\circ$C, 300~$^\circ$C, and 400~$^\circ$C.  
(d) Temperature-dependent Hall effect measured between -5~T and 5~T. The inset indicates a zoomed view of the AHE response at 100 K in a narrower field range of -1~T to 1~T. }
\label{fig:transport}
\end{figure*}

\subsection{Large magneto-transport effects in NCAF films}

The M-H loop in Fig. 3(a) confirms the antiferromagnetic behavior of the Mn$_3$Pt thin films. The observed saturation magnetization of $\sim$ 10 emu/cm$^{3}$ is two orders smaller compared to that of the ferromagnetic Py films (Fig. 3(c)). This weak magnetism arises from the small out-of-plane projection of the Mn spins. Owing to the strong spin--orbit coupling of the Pt atoms, the three Mn moments canted by approximately $0.1^\circ$ out of the Kagome (111) plane \cite{chen2014anomalous}. The small hysteresis is due to the negligible domain reorientation in cubic compounds. The magnetic anisotropy in cubic compounds is almost an order higher than the hexagonal compounds, as the triangular spin configuration point along the three equivalent directions  \cite{zuniga2023observation,sinha2025occurrence}. This is also evident in the reversible M-T curves shown in Fig.~3(b).

Hexagonal members of the family show large bifurcation in the ZFC-FC curves due to negligible magnetic anisotropy in the Kagome plane \cite{sinha2025anomalous}.
Despite a negligibly small net moment, Mn\textsubscript{3}Pt film shows large anomalous Hall resistivity ($\rho_{xy}^{AHE}$) between 2-300 K, as shown in Fig. 3(d). AHE effect in cubic compounds has both intrinsic and extrinsic contributions.  The intrinsic contribution to the anomalous Hall effect (AHE) can be described by the vector spin chirality $\boldsymbol{\epsilon} = \mathbf{S}_1 \times \mathbf{S}_2 + \mathbf{S}_2 \times \mathbf{S}_3 + \mathbf{S}_3 \times \mathbf{S}_1$, whereas the extrinsic contribution arises from the scalar spin chirality $\chi = \mathbf{S}_1 \cdot (\mathbf{S}_2 \times \mathbf{S}_3)$. 
As observed in previous studies, the intrinsic contribution to the AHE is governed by vector spin chirality, which dominates the scattering process \cite{xu2024universal,sinha2025occurrence}.  In cubic compounds, a combination of time-reversal operation and mirror symmetry constitutes a good symmetry. The anomalous Hall effect (AHE) arises as a combined consequence of mirror-symmetry breaking induced by spin canting and the application of a magnetic field. A sizable AHE response has been estimated through Brillouin-zone integration of the Berry curvature component along (111) direction \cite{gurung2019anomalous,chen2014anomalous}. AHE based hysteresis loop indicates a coercivity of $\approx100{mT}$ (inset of Fig. 3(d)). The extraction of $\rho_{xy}^{AHE}$ from $\rho_{xy}$ is discussed in the supplementary section Fig.(S2) \cite{zzSI}. 
The NCAFs with a small coercivity are ideal for the spintronic application, as they allow easy manipulation of the magnetic ordering and also provide a strong readout signal and negligible stray field that enables dense packing.

\begin{figure*}
    \centering
    \includegraphics[width=1\linewidth]{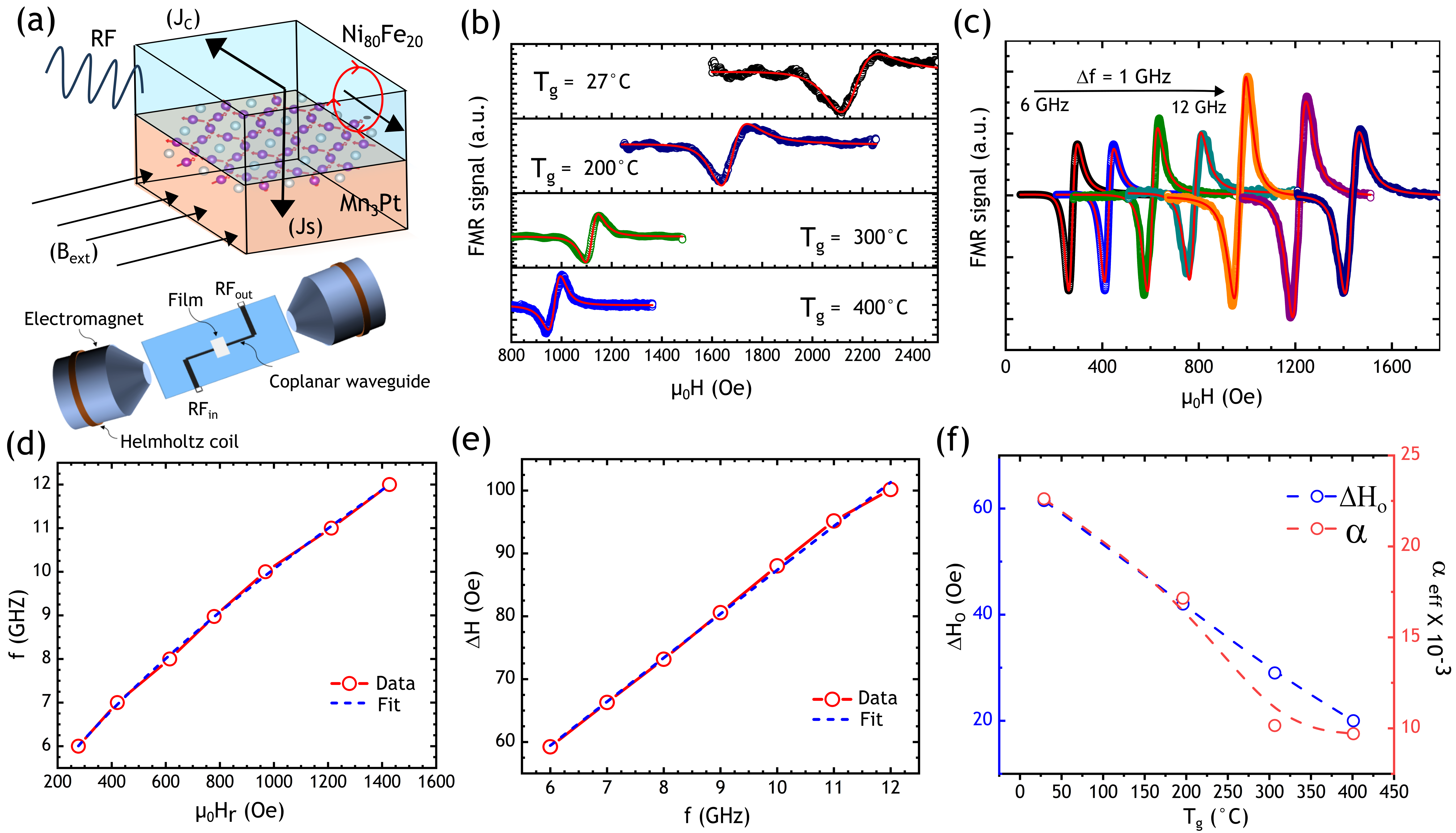}
    \caption{(a) Schematic shows Py/Mn$_{3}$Pt bilayer with direction of microwave field (RF), charge current ($J_{c}$) and spin current ($J_{s}$) for ferromagnetic resonance (FMR) measurements setup. (b) FMR signal obtained at room temperature as a function of external magnetic field  parallel to the film at 10 GHz and its Lorentzian fit obtained for Py samples with growth temperatures (27~$^\circ\text{C}$, 200~$^\circ\text{C}$, 300~$^\circ\text{C}$, 400~$^\circ\text{C}$). The symbols represent experimental data while the lines are the fits. We clearly see reduction in resonance field and sharp spectrum with reduecd FMR linewidth at 400~$^\circ\text{C}$. (c) Typical FMR spectra recorded for Py (10 nm) films (T$_{g}$: 400~$^\circ\text{C}$) for wide frequency range of 6-12 GHz. (d) Frequency versus applied magnetic field plotted and fitted with Kittel equation. (e) FMR linewidth (the full-width at half-maximum of the FMR spectra) plotted vs frequency. (f) Inhomogeneous broadening (10 GHz) and effective Gilbert damping parameter plotted as a function of growth temperature. Open symbols are experimental data while the solid line is fitted data.}
    \label{fig:FMR1}
\end{figure*}

\subsection{Optimization of Py films for FMR}

 To investigate the influence of deposition temperature on magnetic properties, M-H measurements were conducted on  Py (10 nm) thin films deposited at four representative substrate temperatures: 27~$^\circ$C, 200~$^\circ$C, 300~$^\circ$C, and 400~$^\circ$C. The measurement was carried out with a field applied parallel to the surface of the films deposited at four different temperatures, as illustrated in Fig. 3(c). M–H loops reveal that the saturation magnetization ($M_\mathrm{s}$) of Py thin films increases with the growth temperature, reaching a peak at 300~\,$^\circ$C. The maximum $M_\mathrm{s}$ value of 844 $\pm$ 11 emu/cm\textsuperscript{3} is obtained at 300~\,$^\circ$C, beyond which $M_\mathrm{s}$ doesn't show any significant change with temperature.  The initial enhancement in $M_\mathrm{s}$ with temperature is attributed to the crystallization of Py from its amorphous state \cite{dong2021low}.
 \\
 The FMR spectra acquired on  Py thin films grown at different temperatures are shown in  Fig.~4(b). The transmission spectra  measured at frequency 10 GHz for the four representative growth temperatures. The red lines are best fit to the symmetric and antisymmetric Lorentzian functions \cite{Durrenfeld2015} shown in Eq. (1),
\begin{align}
\frac{dP}{dH} =
&- A_{\text{Sym}} \frac{4\Delta H (H - H_{\text{r}})}
{\left[4(H - H_{\text{r}})^2 + (\Delta H)^2 \right]^2} \nonumber \\
&+ A_{\text{Asym}} \frac{(\Delta H)^2 - 4(H - H_{\text{r}})^2}
{\left[4(H - H_{\text{r}})^2 + (\Delta H)^2 \right]^2}
\end{align}
where $A_{\text{Sym}}$ and $A_{\text{Asym}}$ represent the symmetric and antisymmetric absorption coefficients of the Lorentzian function, respectively. H, $H_r$, and $\Delta H$ represent the in-plane external DC magnetic field, resonance field and resonance linewidth (full width at half maximum), respectively. Py films exhibit a clear trend of reduced ferromagnetic resonance (FMR) linewidth broadening ($\Delta H$) and resonance field ($H_r$) with increasing growth temperature, which can be seen in Fig. 4(b). The decrease in $H_r$ from 2100 Oe (27~\,$^\circ$C) to 1000 Oe (400~\,$^\circ$C) is consistent with the enhancement in $M_\mathrm{s}$ with growth temperature.  

This behavior suggests an improvement in the magnetic damping and overall film crystallinity at higher growth temperature (i.e., 400~$^\circ\text{C}$). Fig. 4(c) shows FMR spectra of Py film grown at 400~$^\circ\text{C}$. The measured $H_r$ versus frequency is fitted to the Kittel equation as show in Fig. 4(d) \cite{Jungfleisch2015}:
\begin{equation}
f = \frac{\gamma\mu_{0}}{2\pi} \sqrt{(H_{r} + H_{K}) (H_{r} + 4\pi M_{eff} + H_{K})}
\end{equation}
where $H_{K}$, $\mu_{0}$ and $\gamma$ are surface anisotropy, permeability of free space, and gyromagnetic ratio, respectively.
The effective magnetization $M_{eff}$ obtained from the fitting is close to 740 $emu/cm^3$. The slight variation from squid measurement could be driven by surface anisotropy. 
\begin{figure*}
    \centering
    \includegraphics[width=1\linewidth]{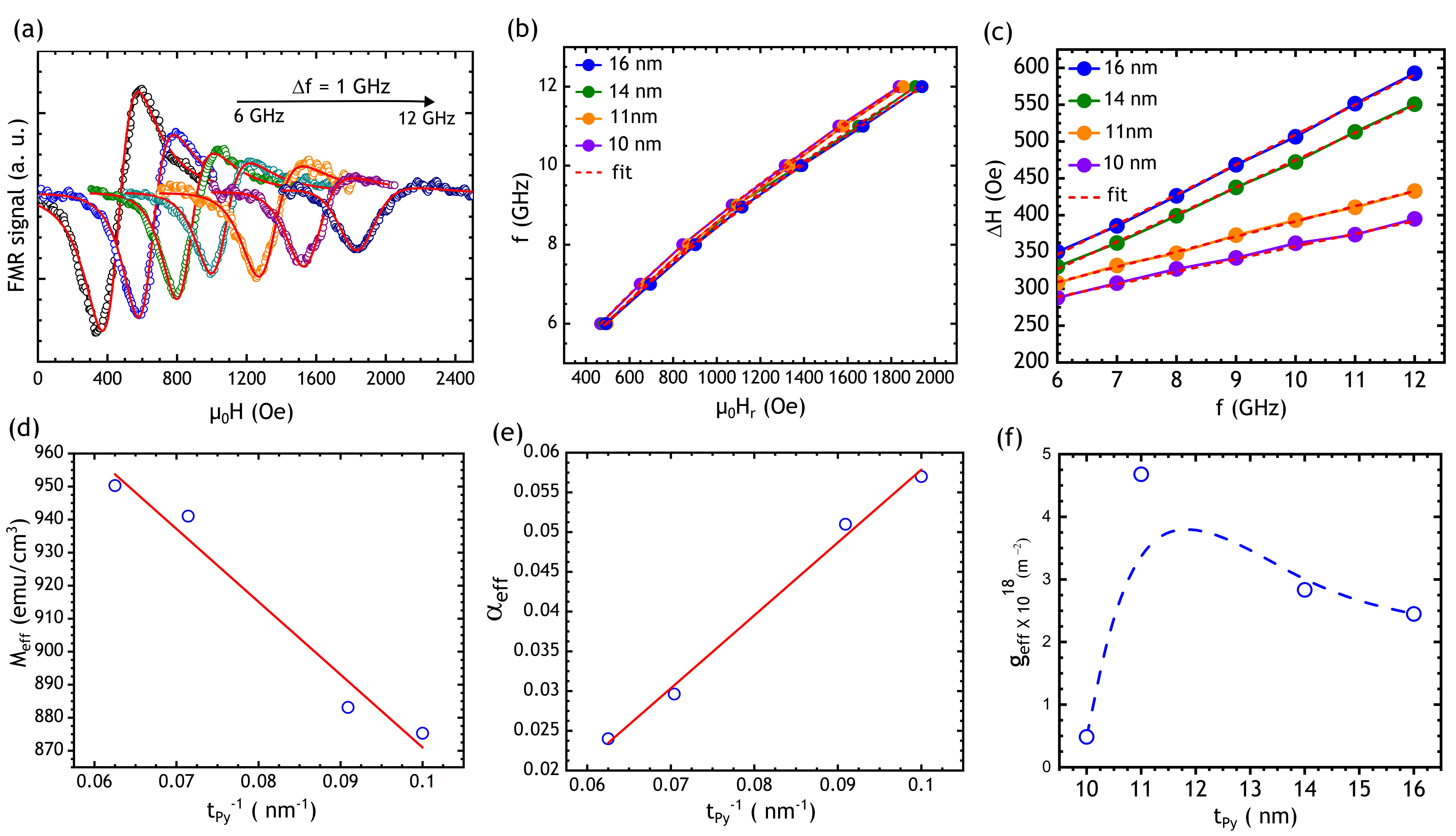}
    \caption{(a) Frequency dependence of FMR spectra obtained for Py(11 nm)/Mn$_{3}$Pt(20 nm). (b)  Frequency (f) plotted as a function of resonance field ($H_{r}$) and fitted with Kittel equation. The thickness of Py film was changed from 10 nm to 16 nm, while keeping Mn$_{3}$Pt thickness constant at 20 nm. (c) The variation of linewidth with frequency (f) and fitted with Eqn(3). (d) Variation of effective magnetization (M$_{eff}$) with inverse thickness ($t_{Py}^{-1}$) and fitted with Eqn(4). (e) Effective Gilbert damping parameter ($\alpha_{eff}$) plotted as a function of inverse thickness ($t_{Py}^{-1}$) and fitted with Eqn 5. (f) The extracted values of spin-mixing conductance (g$^{\uparrow\downarrow}_{eff}$) using Eqn(6) is plotted as a function of thickness.}
    \label{fig:FMR2}
\end{figure*}

FMR linewidth obtained from lorentzian fitting of the measured FMR spectra are plotted as a function of the applied frequency in Fig. 4(e). From the linear fit of the $\Delta H$-f data using the Landau–Lifshitz–Gilbert (LLG) formalism, the effective Gilbert damping parameter ($\alpha_{\mathrm{eff}}$) is extracted from the slope, according to the equation \cite{Sun2020}:
\begin{equation}
\Delta H = \Delta H_0 + \frac{2 \pi \alpha_{\text{eff}} \hbar}{\gamma} f
\end{equation}
where \( \Delta H_0 \) represents the inhomogeneous broadening in FMR linewidth likely arising from the magnetic inhomogeneity due to defects in the sample. Gilbert damping constant ($\alpha_{eff}$) decreased with increasing growth temperature and has the lowest value of 0.00978 in Py films grown at 400$^\circ$C (shown in Fig. 4(f)), which is comparable to previously reported values \cite{Nibarger2003, Hazra2019, Inaba2006}. 

 Between 300~$^{\circ}$C and 400~$^{\circ}$C, the effective damping ($\alpha_{\text{eff}}$) tends to saturate, which is consistent with previous reports on the influence of thermal annealing on Gilbert damping. Beyond a certain threshold temperature (400~$^{\circ}$C),  structural distortions arise due to perturbations in the atomic distribution of the Py layer. This may drive interdiffusion across the interface, as Ni and Fe atoms migrate in opposite directions \cite{schulz2021increase}. The depth profiles obtained from TOF-SIMS reveal no detectable signatures of Ni and Fe ion migration across the interface at 400~$^{\circ}$C.  Additionally \( \Delta H_0 \) is found to decrease with growth temperature. This originated from the inhomogeneities caused by irregular grain boundaries at low-temperature growth. These effects arise due to the local variation of the anisotropy fields in inhomogeneous thin films with  discontinous grains.
Given these findings, we have chosen Py films grown at 400~$^\circ$C for integration into magnetic heterostructures with Mn$_3$Pt films for efficient response of spin transport in antiferromagnetic-ferromagnetic systems \cite{Manschot2004}. 

\subsection{Spin pumping in Py/Mn$_{3}$Pt heterostructure}
FMR spectra measured on Py(11 nm)/Mn$_{3}$Pt(20 nm) bilayers are shown in Fig. 5(a). The field sweeps are measured with an interval of 1 GHz , between 6--12 GHz. The red lines represent the best fit to the data using symmetric and antisymmetric Lorentzian functions, as described by Eq. (1). The variation of f is plotted as a function of ($H_r$) in Fig. 5(b) for all Py thicknesses, and then fitted to the Kittel equation. 

Using these fits, we extract the uniaxial anisotropy field ($H_k$) and effective saturation magnetization ($M_{eff}$) of the ferromagnet. The average value of $H_k$ measured on the four different thicknesses is around 140 Oe. The value of the uniaxial magnetic anisotropy constant ($K_u$) estimated using the relation $K_u = \frac{1}{2} H_K M_s$, is around $5.27 \times 10^4\ \text{erg/cm}^3$. In Fig. 5(d) , the variation of $M_{eff}$ is plotted as a function of thickness and fitted with the equation: 
 \begin{equation}
M_{\text{eff}} = M_S - \frac{2K_S}{\mu_0 M_S} \times t_{\mathrm{Py}}^{-1},
\end{equation}
Where $\mu_{0}$ is the permeability of free space, and $M_\mathrm{s}$ and $K_S$ are the saturation magnetization and surface anisotropy constant, respectively. Based on the fitting, $M_\mathrm{s}$ and $K_S$ are evaluated to be 1089 emu/cm$^{3}$ and 0.42 erg/cm$^{2}$. The obtained saturation magnetization ($M_\mathrm{s}$) values show good agreement between the SQUID data and the FMR measurements.

The variation of $\Delta H$ with frequency for all thicknesses is shown in Fig. 5(c) and fitted using the Landau-Lifshitz-Gilbert (LLG) equation [Eq. (3)]. The linear dependence of $\Delta$H with frequency indicates that the damping in this system is driven by Gilbert type damping. 

The estimated values of ($\alpha_{eff}$) show a monotonic increase with t$^{-1}$ as shown in Fig. 5(e). Interfacing Mn$_3$Pt with Py enhances the Gilbert damping parameter ($\alpha_{eff}$) by one order higher compared to the reference Py films. The enhanced Gilbert damping observed in Py(111)/Mn$_3$Pt(111) is due to spin pumping into the antiferromagnetic Mn$_3$Pt layer, where the injected spin current is absorbed through multiple mechanisms. The uncompensated interfacial spins pinned by antiferromagnetic domains within Mn$_3$Pt, provide a channel for angular momentum transfer from the ferromagnet. Second, strong bulk spin–orbit coupling in Mn$_3$Pt relaxes the pumped spin angular momentum, acting as a facet-independent mechanism that contributes to damping. Third, the noncollinear magnetic structure of Mn$_3$Pt allows for the emergence of the magnetic spin Hall effect (MSHE), which can further mediate spin current absorption depending on current and field orientation \cite{Holanda2020,lund2021spin,frangou2016enhanced, Pal2024}. 
The noncollinear magnetic order of Mn$_3$Pt provides spin channels for transverse spin absorption, enhancing the damping and leading to larger linewidth broadening in FMR.

$\alpha_{eff}$ is linearly fitted to the inverse of thickness $t_{py}^{-1}$ using the following expression \cite{khan2024magnetodynamic}:
\begin{equation}
\alpha_{\text{eff}} = \alpha_{\text{int}} + g_{\text{eff}}^{\uparrow\downarrow} \frac{\gamma \hbar}{4\pi M_S} \times t_{\mathrm{py}}^{-1}
\end{equation}
where $\alpha_{\text{int}}$ represents the intrinsic Gilbert damping constant and $g_{\text{eff}}^{\uparrow\downarrow}$ denotes the effective spin mixing conductance of Py/Mn$_{3}$Pt heterostructures. The gyromagnetic motion of the spins in the Py layer exerts a momentum across the interface that is explained by the parameter called the  spin-mixing conductance ($g_{\text{eff}}^{\uparrow\downarrow}$) \cite{ding2024mitigation}. Based on the fitting parameters, the values of $\alpha_{\text{in}}$ and $g_{\text{eff}}^{\uparrow\downarrow}$ are $3.1  \times 10^{-2}$ and $4.8  \times 10^{18}\,\mathrm{m}^{-2}$, respectively, the order of which is comparable to other Py/Mn$_{3}$X/Pt heterostructures \cite{Holanda2020,mosendz2010quantifying,kimata2019magnetic,hayashi2021spin}. 
The increase in damping parameter as a consequence of enhanced spin pumping is related to the effective spin mixing conductance (g$^{\uparrow\downarrow}_{eff}$) and thickness of the Py film ($t_{Ni_{80}Fe_{20}}$), which is given by the following expression \cite{Wang2014}:
\begin{equation}
g_{\text{eff}}^{\uparrow\downarrow} = \frac{M_{\text{eff}} 
 t_{\text{Ni}_{80}\text{Fe}_{20}}}{g \mu_B} \left( \alpha_{\text{Ni}_{80}\text{Fe}_{20}/\text{Mn}_3\text{Pt}} - \alpha_{\text{Ni}_{80}\text{Fe}_{20}} \right)
\end{equation}
In this equation, $\left( \alpha_{\text{Ni}_{80}\text{Fe}_{20}/\text{Mn}_3\text{Pt}} - \alpha_{\text{Ni}_{80}\text{Fe}_{20}} \right)$ is the damping enhancement by spin pumping. The real part of the effective spin-mixing conductance, g$^{\uparrow\downarrow}_{eff}$, governs spin angular momentum transfer across the interface, while the imaginary part is typically negligible.

The plot of g$^{\uparrow\downarrow}_{eff}$ with thickness of Py layer shows a non-monotonic behavior, which increases up to $\sim$ 11 nm, then decreases or saturates. The behavior of g$^{\uparrow\downarrow}_{eff}$ with Py thickness reflects the dynamic balance between efficient spin transfer and loss mechanisms like spin backflow and interface scattering \cite{kumar2025effect, sharma2022magnetization}. The peak indicates an optimal thickness for maximizing spin pumping efficiency at the interface. These results together provide strong evidence for spin pumping across the Py/Mn$_3$Pt interface. This work shows that chiral antiferromagnetic systems, such as Mn$_3$Pt can offer an alternative gateway to manifest spin to charge degrees of freedom through the magnetic inverse spin Hall effect.

\section{Conclusion}
In this work, we establish Mn$_3$Pt, a noncollinear antiferromagnet with strong Berry-curvature, as an efficient spin sink in Py/Mn$_3$Pt heterostructures. Chemical and structural analyses down to atomic scale resolution confirm atomically sharp and low-roughness interfaces. While the optimized Py layers with intrinsically low damping provide a robust platform for precise ferromagnetic resonance (FMR) studies, Mn$_3$Pt thin films exhibit a pronounced anomalous Hall effect (AHE) with a low coercivity, reflecting the Berry-curvature-driven transport response associated with the nontrivial topology of their band structure. When interfaced with Py, the FMR spectra display a significant linewidth broadening compared to reference Py films. This enhanced damping originates from spin pumping across the interface, where precessing spins in Py efficiently transfer angular momentum into the Mn$_3$Pt layer. The noncollinear magnetic order of Mn$_3$Pt acts as an effective spin-absorption channel, suppressing backflow and reinforcing interfacial spin transparency.  The combination of a strong AHE response and broadened FMR spectra demonstrates the dual role of Mn$_3$Pt, as a topological material enabling robust charge-spin conversion and as a highly efficient sink for transverse spin currents. These findings highlight Mn$_3$Pt based bilayers as a promising platform for next-generation antiferromagnetic spintronic devices, offering tunable damping, high spin-mixing conductance, and low-power operation.  
\section{ACKNOWLEDGMENTS}
This work was suppoerted by ANRF Core Research Grant (CRG/2023/008193), Government of India. IS and SM acknowledge partial support from the DST-funded project 'CONCEPT' under the National Programme on Nano Science and Technology (DST/NM/QM-10/2019), Government of India.  We acknowledge the Department of Physics at IIT Delhi for XRD, PPMS, and MPMS measurements facility. The authors also thank the Central Research Facility (IIT Delhi) for TOF-SIMS depth profiling measurements.

The authors declare no competing financial interests.

\bibliographystyle{apsrev4-2}
\bibliography{references}

\begin{thebibliography}{59}%
\makeatletter
\providecommand \@ifxundefined [1]{%
 \@ifx{#1\undefined}
}%
\providecommand \@ifnum [1]{%
 \ifnum #1\expandafter \@firstoftwo
 \else \expandafter \@secondoftwo
 \fi
}%
\providecommand \@ifx [1]{%
 \ifx #1\expandafter \@firstoftwo
 \else \expandafter \@secondoftwo
 \fi
}%
\providecommand \natexlab [1]{#1}%
\providecommand \enquote  [1]{``#1''}%
\providecommand \bibnamefont  [1]{#1}%
\providecommand \bibfnamefont [1]{#1}%
\providecommand \citenamefont [1]{#1}%
\providecommand \href@noop [0]{\@secondoftwo}%
\providecommand \href [0]{\begingroup \@sanitize@url \@href}%
\providecommand \@href[1]{\@@startlink{#1}\@@href}%
\providecommand \@@href[1]{\endgroup#1\@@endlink}%
\providecommand \@sanitize@url [0]{\catcode `\\12\catcode `\$12\catcode `\&12\catcode `\#12\catcode `\^12\catcode `\_12\catcode `\%12\relax}%
\providecommand \@@startlink[1]{}%
\providecommand \@@endlink[0]{}%
\providecommand \url  [0]{\begingroup\@sanitize@url \@url }%
\providecommand \@url [1]{\endgroup\@href {#1}{\urlprefix }}%
\providecommand \urlprefix  [0]{URL }%
\providecommand \Eprint [0]{\href }%
\providecommand \doibase [0]{https://doi.org/}%
\providecommand \selectlanguage [0]{\@gobble}%
\providecommand \bibinfo  [0]{\@secondoftwo}%
\providecommand \bibfield  [0]{\@secondoftwo}%
\providecommand \translation [1]{[#1]}%
\providecommand \BibitemOpen [0]{}%
\providecommand \bibitemStop [0]{}%
\providecommand \bibitemNoStop [0]{.\EOS\space}%
\providecommand \EOS [0]{\spacefactor3000\relax}%
\providecommand \BibitemShut  [1]{\csname bibitem#1\endcsname}%
\let\auto@bib@innerbib\@empty
\bibitem [{\citenamefont {Han}\ \emph {et~al.}(2018)\citenamefont {Han}, \citenamefont {Otani},\ and\ \citenamefont {Maekawa}}]{Han2018}%
  \BibitemOpen
  \bibfield  {author} {\bibinfo {author} {\bibfnamefont {W.}~\bibnamefont {Han}}, \bibinfo {author} {\bibfnamefont {Y.}~\bibnamefont {Otani}},\ and\ \bibinfo {author} {\bibfnamefont {S.}~\bibnamefont {Maekawa}},\ }\href@noop {} {\bibfield  {journal} {\bibinfo  {journal} {npj Quantum Materials}\ }\textbf {\bibinfo {volume} {3}},\ \bibinfo {pages} {27} (\bibinfo {year} {2018})}\BibitemShut {NoStop}%
\bibitem [{\citenamefont {Mangin}\ \emph {et~al.}(2006)\citenamefont {Mangin}, \citenamefont {Ravelosona}, \citenamefont {Katine}, \citenamefont {Carey}, \citenamefont {Terris},\ and\ \citenamefont {Fullerton}}]{Mangin2006}%
  \BibitemOpen
  \bibfield  {author} {\bibinfo {author} {\bibfnamefont {S.}~\bibnamefont {Mangin}}, \bibinfo {author} {\bibfnamefont {D.}~\bibnamefont {Ravelosona}}, \bibinfo {author} {\bibfnamefont {J.~A.}\ \bibnamefont {Katine}}, \bibinfo {author} {\bibfnamefont {M.~J.}\ \bibnamefont {Carey}}, \bibinfo {author} {\bibfnamefont {B.~D.}\ \bibnamefont {Terris}},\ and\ \bibinfo {author} {\bibfnamefont {E.~E.}\ \bibnamefont {Fullerton}},\ }\href@noop {} {\bibfield  {journal} {\bibinfo  {journal} {Nature Materials}\ }\textbf {\bibinfo {volume} {5}},\ \bibinfo {pages} {210} (\bibinfo {year} {2006})}\BibitemShut {NoStop}%
\bibitem [{\citenamefont {Chowdhury}\ \emph {et~al.}(2023)\citenamefont {Chowdhury}, \citenamefont {Khan}, \citenamefont {Bangar}, \citenamefont {Gupta}, \citenamefont {Yadav}, \citenamefont {Agarwal}, \citenamefont {Kumar},\ and\ \citenamefont {Muduli}}]{Chowdhury2023}%
  \BibitemOpen
  \bibfield  {author} {\bibinfo {author} {\bibfnamefont {N.}~\bibnamefont {Chowdhury}}, \bibinfo {author} {\bibfnamefont {K.~I.~A.}\ \bibnamefont {Khan}}, \bibinfo {author} {\bibfnamefont {H.}~\bibnamefont {Bangar}}, \bibinfo {author} {\bibfnamefont {P.}~\bibnamefont {Gupta}}, \bibinfo {author} {\bibfnamefont {R.~S.}\ \bibnamefont {Yadav}}, \bibinfo {author} {\bibfnamefont {R.}~\bibnamefont {Agarwal}}, \bibinfo {author} {\bibfnamefont {A.}~\bibnamefont {Kumar}},\ and\ \bibinfo {author} {\bibfnamefont {P.~K.}\ \bibnamefont {Muduli}},\ }\href@noop {} {\bibfield  {journal} {\bibinfo  {journal} {Proceedings of the National Academy of Sciences, India Section A: Physical Sciences}\ }\textbf {\bibinfo {volume} {93}},\ \bibinfo {pages} {477} (\bibinfo {year} {2023})}\BibitemShut {NoStop}%
\bibitem [{\citenamefont {Bai}\ \emph {et~al.}(2022)\citenamefont {Bai}, \citenamefont {Zhang}, \citenamefont {Han}, \citenamefont {Zhou}, \citenamefont {Pan},\ and\ \citenamefont {Song}}]{bai2022antiferromagnetism}%
  \BibitemOpen
  \bibfield  {author} {\bibinfo {author} {\bibfnamefont {H.}~\bibnamefont {Bai}}, \bibinfo {author} {\bibfnamefont {Y.}~\bibnamefont {Zhang}}, \bibinfo {author} {\bibfnamefont {L.}~\bibnamefont {Han}}, \bibinfo {author} {\bibfnamefont {Y.}~\bibnamefont {Zhou}}, \bibinfo {author} {\bibfnamefont {F.}~\bibnamefont {Pan}},\ and\ \bibinfo {author} {\bibfnamefont {C.}~\bibnamefont {Song}},\ }\href@noop {} {\bibfield  {journal} {\bibinfo  {journal} {Applied Physics Reviews}\ }\textbf {\bibinfo {volume} {9}} (\bibinfo {year} {2022})}\BibitemShut {NoStop}%
\bibitem [{\citenamefont {Shiino}\ \emph {et~al.}(2016)\citenamefont {Shiino}, \citenamefont {Oh}, \citenamefont {Haney}, \citenamefont {Lee}, \citenamefont {Go}, \citenamefont {Park},\ and\ \citenamefont {Lee}}]{shiino2016antiferromagnetic}%
  \BibitemOpen
  \bibfield  {author} {\bibinfo {author} {\bibfnamefont {T.}~\bibnamefont {Shiino}}, \bibinfo {author} {\bibfnamefont {S.-H.}\ \bibnamefont {Oh}}, \bibinfo {author} {\bibfnamefont {P.~M.}\ \bibnamefont {Haney}}, \bibinfo {author} {\bibfnamefont {S.-W.}\ \bibnamefont {Lee}}, \bibinfo {author} {\bibfnamefont {G.}~\bibnamefont {Go}}, \bibinfo {author} {\bibfnamefont {B.-G.}\ \bibnamefont {Park}},\ and\ \bibinfo {author} {\bibfnamefont {K.-J.}\ \bibnamefont {Lee}},\ }\href@noop {} {\bibfield  {journal} {\bibinfo  {journal} {Physical Review Letters}\ }\textbf {\bibinfo {volume} {117}},\ \bibinfo {pages} {087203} (\bibinfo {year} {2016})}\BibitemShut {NoStop}%
\bibitem [{\citenamefont {Wu}\ \emph {et~al.}(2024)\citenamefont {Wu}, \citenamefont {Chen}, \citenamefont {Nomoto}, \citenamefont {Tserkovnyak}, \citenamefont {Isshiki}, \citenamefont {Nakatani}, \citenamefont {Higo}, \citenamefont {Tomita}, \citenamefont {Kondou}, \citenamefont {Arita} \emph {et~al.}}]{wu2024current}%
  \BibitemOpen
  \bibfield  {author} {\bibinfo {author} {\bibfnamefont {M.}~\bibnamefont {Wu}}, \bibinfo {author} {\bibfnamefont {T.}~\bibnamefont {Chen}}, \bibinfo {author} {\bibfnamefont {T.}~\bibnamefont {Nomoto}}, \bibinfo {author} {\bibfnamefont {Y.}~\bibnamefont {Tserkovnyak}}, \bibinfo {author} {\bibfnamefont {H.}~\bibnamefont {Isshiki}}, \bibinfo {author} {\bibfnamefont {Y.}~\bibnamefont {Nakatani}}, \bibinfo {author} {\bibfnamefont {T.}~\bibnamefont {Higo}}, \bibinfo {author} {\bibfnamefont {T.}~\bibnamefont {Tomita}}, \bibinfo {author} {\bibfnamefont {K.}~\bibnamefont {Kondou}}, \bibinfo {author} {\bibfnamefont {R.}~\bibnamefont {Arita}}, \emph {et~al.},\ }\href@noop {} {\bibfield  {journal} {\bibinfo  {journal} {Nature Communications}\ }\textbf {\bibinfo {volume} {15}},\ \bibinfo {pages} {4305} (\bibinfo {year} {2024})}\BibitemShut {NoStop}%
\bibitem [{\citenamefont {Nakatsuji}\ \emph {et~al.}(2015)\citenamefont {Nakatsuji}, \citenamefont {Kiyohara},\ and\ \citenamefont {Higo}}]{NCAFM_AHE_Nat_Nakatsuji2015}%
  \BibitemOpen
  \bibfield  {author} {\bibinfo {author} {\bibfnamefont {S.}~\bibnamefont {Nakatsuji}}, \bibinfo {author} {\bibfnamefont {N.}~\bibnamefont {Kiyohara}},\ and\ \bibinfo {author} {\bibfnamefont {T.}~\bibnamefont {Higo}},\ }\href@noop {} {\bibfield  {journal} {\bibinfo  {journal} {Nat.}\ }\textbf {\bibinfo {volume} {527}},\ \bibinfo {pages} {212–215} (\bibinfo {year} {2015})}\BibitemShut {NoStop}%
\bibitem [{\citenamefont {Ikhlas}\ \emph {et~al.}(2017)\citenamefont {Ikhlas}, \citenamefont {Tomita}, \citenamefont {Koretsune}, \citenamefont {Suzuki}, \citenamefont {Nishio-Hamane}, \citenamefont {Arita}, \citenamefont {Otani},\ and\ \citenamefont {Nakatsuji}}]{NCAFM_ANE_Nat_phy_Ikhlas2017}%
  \BibitemOpen
  \bibfield  {author} {\bibinfo {author} {\bibfnamefont {M.}~\bibnamefont {Ikhlas}}, \bibinfo {author} {\bibfnamefont {T.}~\bibnamefont {Tomita}}, \bibinfo {author} {\bibfnamefont {T.}~\bibnamefont {Koretsune}}, \bibinfo {author} {\bibfnamefont {M.-T.}\ \bibnamefont {Suzuki}}, \bibinfo {author} {\bibfnamefont {D.}~\bibnamefont {Nishio-Hamane}}, \bibinfo {author} {\bibfnamefont {R.}~\bibnamefont {Arita}}, \bibinfo {author} {\bibfnamefont {Y.}~\bibnamefont {Otani}},\ and\ \bibinfo {author} {\bibfnamefont {S.}~\bibnamefont {Nakatsuji}},\ }\href@noop {} {\bibfield  {journal} {\bibinfo  {journal} {Nat. Phys.}\ }\textbf {\bibinfo {volume} {13}},\ \bibinfo {pages} {1085–1090} (\bibinfo {year} {2017})}\BibitemShut {NoStop}%
\bibitem [{\citenamefont {Pandey}\ \emph {et~al.}(2024)\citenamefont {Pandey}, \citenamefont {Deka}, \citenamefont {Yoon}, \citenamefont {Mathew}, \citenamefont {Koerner}, \citenamefont {Dreyer}, \citenamefont {Taylor}, \citenamefont {Parkin},\ and\ \citenamefont {Woltersdorf}}]{pandey_ane}%
  \BibitemOpen
  \bibfield  {author} {\bibinfo {author} {\bibfnamefont {A.}~\bibnamefont {Pandey}}, \bibinfo {author} {\bibfnamefont {J.}~\bibnamefont {Deka}}, \bibinfo {author} {\bibfnamefont {J.}~\bibnamefont {Yoon}}, \bibinfo {author} {\bibfnamefont {A.}~\bibnamefont {Mathew}}, \bibinfo {author} {\bibfnamefont {C.}~\bibnamefont {Koerner}}, \bibinfo {author} {\bibfnamefont {R.}~\bibnamefont {Dreyer}}, \bibinfo {author} {\bibfnamefont {J.~M.}\ \bibnamefont {Taylor}}, \bibinfo {author} {\bibfnamefont {S.~S.~P.}\ \bibnamefont {Parkin}},\ and\ \bibinfo {author} {\bibfnamefont {G.}~\bibnamefont {Woltersdorf}},\ }\href@noop {} {\bibfield  {journal} {\bibinfo  {journal} {ACS Nano}\ }\textbf {\bibinfo {volume} {18}},\ \bibinfo {pages} {31949} (\bibinfo {year} {2024})}\BibitemShut {NoStop}%
\bibitem [{\citenamefont {Higo}\ \emph {et~al.}(2018)\citenamefont {Higo}, \citenamefont {Man}, \citenamefont {Gopman}, \citenamefont {Wu}, \citenamefont {Koretsune}, \citenamefont {van~’t Erve}, \citenamefont {Kabanov}, \citenamefont {Rees}, \citenamefont {Li}, \citenamefont {Suzuki}, \citenamefont {Patankar}, \citenamefont {Ikhlas}, \citenamefont {Chien}, \citenamefont {Arita}, \citenamefont {Shull}, \citenamefont {Orenstein},\ and\ \citenamefont {Nakatsuji}}]{NCAFM_MOKE_nat_photo_Higo2018}%
  \BibitemOpen
  \bibfield  {author} {\bibinfo {author} {\bibfnamefont {T.}~\bibnamefont {Higo}}, \bibinfo {author} {\bibfnamefont {H.}~\bibnamefont {Man}}, \bibinfo {author} {\bibfnamefont {D.~B.}\ \bibnamefont {Gopman}}, \bibinfo {author} {\bibfnamefont {L.}~\bibnamefont {Wu}}, \bibinfo {author} {\bibfnamefont {T.}~\bibnamefont {Koretsune}}, \bibinfo {author} {\bibfnamefont {O.~M.~J.}\ \bibnamefont {van~’t Erve}}, \bibinfo {author} {\bibfnamefont {Y.~P.}\ \bibnamefont {Kabanov}}, \bibinfo {author} {\bibfnamefont {D.}~\bibnamefont {Rees}}, \bibinfo {author} {\bibfnamefont {Y.}~\bibnamefont {Li}}, \bibinfo {author} {\bibfnamefont {M.-T.}\ \bibnamefont {Suzuki}}, \bibinfo {author} {\bibfnamefont {S.}~\bibnamefont {Patankar}}, \bibinfo {author} {\bibfnamefont {M.}~\bibnamefont {Ikhlas}}, \bibinfo {author} {\bibfnamefont {C.~L.}\ \bibnamefont {Chien}}, \bibinfo {author} {\bibfnamefont {R.}~\bibnamefont {Arita}}, \bibinfo {author} {\bibfnamefont {R.~D.}\ \bibnamefont {Shull}}, \bibinfo {author} {\bibfnamefont
  {J.}~\bibnamefont {Orenstein}},\ and\ \bibinfo {author} {\bibfnamefont {S.}~\bibnamefont {Nakatsuji}},\ }\href@noop {} {\bibfield  {journal} {\bibinfo  {journal} {Nature Photonics}\ }\textbf {\bibinfo {volume} {12}},\ \bibinfo {pages} {73–78} (\bibinfo {year} {2018})}\BibitemShut {NoStop}%
\bibitem [{\citenamefont {Yang}\ \emph {et~al.}(2017)\citenamefont {Yang}, \citenamefont {Sun}, \citenamefont {Zhang}, \citenamefont {Shi}, \citenamefont {Parkin},\ and\ \citenamefont {Yan}}]{yang2017topological}%
  \BibitemOpen
  \bibfield  {author} {\bibinfo {author} {\bibfnamefont {H.}~\bibnamefont {Yang}}, \bibinfo {author} {\bibfnamefont {Y.}~\bibnamefont {Sun}}, \bibinfo {author} {\bibfnamefont {Y.}~\bibnamefont {Zhang}}, \bibinfo {author} {\bibfnamefont {W.-J.}\ \bibnamefont {Shi}}, \bibinfo {author} {\bibfnamefont {S.~S.}\ \bibnamefont {Parkin}},\ and\ \bibinfo {author} {\bibfnamefont {B.}~\bibnamefont {Yan}},\ }\href@noop {} {\bibfield  {journal} {\bibinfo  {journal} {New Journal of Physics}\ }\textbf {\bibinfo {volume} {19}},\ \bibinfo {pages} {015008} (\bibinfo {year} {2017})}\BibitemShut {NoStop}%
\bibitem [{\citenamefont {Busch}\ \emph {et~al.}(2021)\citenamefont {Busch}, \citenamefont {G{\"o}bel},\ and\ \citenamefont {Mertig}}]{busch2021spin}%
  \BibitemOpen
  \bibfield  {author} {\bibinfo {author} {\bibfnamefont {O.}~\bibnamefont {Busch}}, \bibinfo {author} {\bibfnamefont {B.}~\bibnamefont {G{\"o}bel}},\ and\ \bibinfo {author} {\bibfnamefont {I.}~\bibnamefont {Mertig}},\ }\href@noop {} {\bibfield  {journal} {\bibinfo  {journal} {Physical Review B}\ }\textbf {\bibinfo {volume} {104}},\ \bibinfo {pages} {184423} (\bibinfo {year} {2021})}\BibitemShut {NoStop}%
\bibitem [{\citenamefont {Wu}\ \emph {et~al.}(2020)\citenamefont {Wu}, \citenamefont {Isshiki}, \citenamefont {Chen}, \citenamefont {Higo}, \citenamefont {Nakatsuji},\ and\ \citenamefont {Otani}}]{wu2020magneto}%
  \BibitemOpen
  \bibfield  {author} {\bibinfo {author} {\bibfnamefont {M.}~\bibnamefont {Wu}}, \bibinfo {author} {\bibfnamefont {H.}~\bibnamefont {Isshiki}}, \bibinfo {author} {\bibfnamefont {T.}~\bibnamefont {Chen}}, \bibinfo {author} {\bibfnamefont {T.}~\bibnamefont {Higo}}, \bibinfo {author} {\bibfnamefont {S.}~\bibnamefont {Nakatsuji}},\ and\ \bibinfo {author} {\bibfnamefont {Y.}~\bibnamefont {Otani}},\ }\href@noop {} {\bibfield  {journal} {\bibinfo  {journal} {Applied Physics Letters}\ }\textbf {\bibinfo {volume} {116}} (\bibinfo {year} {2020})}\BibitemShut {NoStop}%
\bibitem [{\citenamefont {Nayak}\ \emph {et~al.}(2016)\citenamefont {Nayak}, \citenamefont {Fischer}, \citenamefont {Sun}, \citenamefont {Yan}, \citenamefont {Karel}, \citenamefont {Komarek}, \citenamefont {Shekhar}, \citenamefont {Kumar}, \citenamefont {Schnelle}, \citenamefont {Kübler},\ and\ \citenamefont {Felser}}]{Nayak2016}%
  \BibitemOpen
  \bibfield  {author} {\bibinfo {author} {\bibfnamefont {A.~K.}\ \bibnamefont {Nayak}}, \bibinfo {author} {\bibfnamefont {J.~E.}\ \bibnamefont {Fischer}}, \bibinfo {author} {\bibfnamefont {Y.}~\bibnamefont {Sun}}, \bibinfo {author} {\bibfnamefont {B.}~\bibnamefont {Yan}}, \bibinfo {author} {\bibfnamefont {J.}~\bibnamefont {Karel}}, \bibinfo {author} {\bibfnamefont {A.~C.}\ \bibnamefont {Komarek}}, \bibinfo {author} {\bibfnamefont {C.}~\bibnamefont {Shekhar}}, \bibinfo {author} {\bibfnamefont {N.}~\bibnamefont {Kumar}}, \bibinfo {author} {\bibfnamefont {W.}~\bibnamefont {Schnelle}}, \bibinfo {author} {\bibfnamefont {J.}~\bibnamefont {Kübler}},\ and\ \bibinfo {author} {\bibfnamefont {C.}~\bibnamefont {Felser}},\ }\href@noop {} {\bibfield  {journal} {\bibinfo  {journal} {Science Advances}\ }\textbf {\bibinfo {volume} {2}},\ \bibinfo {pages} {e1501870} (\bibinfo {year} {2016})}\BibitemShut {NoStop}%
\bibitem [{\citenamefont {Zuniga-Cespedes}\ \emph {et~al.}(2023)\citenamefont {Zuniga-Cespedes}, \citenamefont {Manna}, \citenamefont {Noad}, \citenamefont {Yang}, \citenamefont {Nicklas}, \citenamefont {Felser}, \citenamefont {Mackenzie},\ and\ \citenamefont {Hicks}}]{zuniga2023observation}%
  \BibitemOpen
  \bibfield  {author} {\bibinfo {author} {\bibfnamefont {B.~E.}\ \bibnamefont {Zuniga-Cespedes}}, \bibinfo {author} {\bibfnamefont {K.}~\bibnamefont {Manna}}, \bibinfo {author} {\bibfnamefont {H.~M.}\ \bibnamefont {Noad}}, \bibinfo {author} {\bibfnamefont {P.-Y.}\ \bibnamefont {Yang}}, \bibinfo {author} {\bibfnamefont {M.}~\bibnamefont {Nicklas}}, \bibinfo {author} {\bibfnamefont {C.}~\bibnamefont {Felser}}, \bibinfo {author} {\bibfnamefont {A.~P.}\ \bibnamefont {Mackenzie}},\ and\ \bibinfo {author} {\bibfnamefont {C.~W.}\ \bibnamefont {Hicks}},\ }\href@noop {} {\bibfield  {journal} {\bibinfo  {journal} {New Journal of Physics}\ }\textbf {\bibinfo {volume} {25}},\ \bibinfo {pages} {023029} (\bibinfo {year} {2023})}\BibitemShut {NoStop}%
\bibitem [{\citenamefont {Li}\ \emph {et~al.}(2023)\citenamefont {Li}, \citenamefont {Koo}, \citenamefont {Zhu}, \citenamefont {Behnia},\ and\ \citenamefont {Yan}}]{li2023field}%
  \BibitemOpen
  \bibfield  {author} {\bibinfo {author} {\bibfnamefont {X.}~\bibnamefont {Li}}, \bibinfo {author} {\bibfnamefont {J.}~\bibnamefont {Koo}}, \bibinfo {author} {\bibfnamefont {Z.}~\bibnamefont {Zhu}}, \bibinfo {author} {\bibfnamefont {K.}~\bibnamefont {Behnia}},\ and\ \bibinfo {author} {\bibfnamefont {B.}~\bibnamefont {Yan}},\ }\href@noop {} {\bibfield  {journal} {\bibinfo  {journal} {Nature Communications}\ }\textbf {\bibinfo {volume} {14}},\ \bibinfo {pages} {1642} (\bibinfo {year} {2023})}\BibitemShut {NoStop}%
\bibitem [{\citenamefont {Nagaosa}\ \emph {et~al.}(2010)\citenamefont {Nagaosa}, \citenamefont {Sinova}, \citenamefont {Onoda}, \citenamefont {MacDonald},\ and\ \citenamefont {Ong}}]{Nagaosa2010}%
  \BibitemOpen
  \bibfield  {author} {\bibinfo {author} {\bibfnamefont {N.}~\bibnamefont {Nagaosa}}, \bibinfo {author} {\bibfnamefont {J.}~\bibnamefont {Sinova}}, \bibinfo {author} {\bibfnamefont {S.}~\bibnamefont {Onoda}}, \bibinfo {author} {\bibfnamefont {A.~H.}\ \bibnamefont {MacDonald}},\ and\ \bibinfo {author} {\bibfnamefont {N.~P.}\ \bibnamefont {Ong}},\ }\href@noop {} {\bibfield  {journal} {\bibinfo  {journal} {Reviews of Modern Physics}\ }\textbf {\bibinfo {volume} {82}},\ \bibinfo {pages} {1539} (\bibinfo {year} {2010})}\BibitemShut {NoStop}%
\bibitem [{\citenamefont {Fukami}\ \emph {et~al.}(2016)\citenamefont {Fukami}, \citenamefont {Zhang}, \citenamefont {DuttaGupta}, \citenamefont {Kurenkov},\ and\ \citenamefont {Ohno}}]{Fukami2016}%
  \BibitemOpen
  \bibfield  {author} {\bibinfo {author} {\bibfnamefont {S.}~\bibnamefont {Fukami}}, \bibinfo {author} {\bibfnamefont {C.}~\bibnamefont {Zhang}}, \bibinfo {author} {\bibfnamefont {S.}~\bibnamefont {DuttaGupta}}, \bibinfo {author} {\bibfnamefont {A.}~\bibnamefont {Kurenkov}},\ and\ \bibinfo {author} {\bibfnamefont {H.}~\bibnamefont {Ohno}},\ }\href@noop {} {\bibfield  {journal} {\bibinfo  {journal} {Nature Materials}\ }\textbf {\bibinfo {volume} {15}},\ \bibinfo {pages} {535} (\bibinfo {year} {2016})}\BibitemShut {NoStop}%
\bibitem [{\citenamefont {Poelchen}\ \emph {et~al.}(2023)\citenamefont {Poelchen}, \citenamefont {Hellwig}, \citenamefont {Peters}, \citenamefont {Usachov}, \citenamefont {Kliemt}, \citenamefont {Laubschat}, \citenamefont {Echenique}, \citenamefont {Chulkov}, \citenamefont {Krellner}, \citenamefont {Parkin} \emph {et~al.}}]{poelchen2023long}%
  \BibitemOpen
  \bibfield  {author} {\bibinfo {author} {\bibfnamefont {G.}~\bibnamefont {Poelchen}}, \bibinfo {author} {\bibfnamefont {J.}~\bibnamefont {Hellwig}}, \bibinfo {author} {\bibfnamefont {M.}~\bibnamefont {Peters}}, \bibinfo {author} {\bibfnamefont {D.~Y.}\ \bibnamefont {Usachov}}, \bibinfo {author} {\bibfnamefont {K.}~\bibnamefont {Kliemt}}, \bibinfo {author} {\bibfnamefont {C.}~\bibnamefont {Laubschat}}, \bibinfo {author} {\bibfnamefont {P.~M.}\ \bibnamefont {Echenique}}, \bibinfo {author} {\bibfnamefont {E.~V.}\ \bibnamefont {Chulkov}}, \bibinfo {author} {\bibfnamefont {C.}~\bibnamefont {Krellner}}, \bibinfo {author} {\bibfnamefont {S.}~\bibnamefont {Parkin}}, \emph {et~al.},\ }\href@noop {} {\bibfield  {journal} {\bibinfo  {journal} {Nature Communications}\ }\textbf {\bibinfo {volume} {14}},\ \bibinfo {pages} {5422} (\bibinfo {year} {2023})}\BibitemShut {NoStop}%
\bibitem [{\citenamefont {Pal}\ \emph {et~al.}(2022)\citenamefont {Pal}, \citenamefont {Hazra}, \citenamefont {G{\"o}bel}, \citenamefont {Jeon}, \citenamefont {Pandeya}, \citenamefont {Chakraborty}, \citenamefont {Busch}, \citenamefont {Srivastava}, \citenamefont {Deniz}, \citenamefont {Taylor} \emph {et~al.}}]{pal2022setting}%
  \BibitemOpen
  \bibfield  {author} {\bibinfo {author} {\bibfnamefont {B.}~\bibnamefont {Pal}}, \bibinfo {author} {\bibfnamefont {B.~K.}\ \bibnamefont {Hazra}}, \bibinfo {author} {\bibfnamefont {B.}~\bibnamefont {G{\"o}bel}}, \bibinfo {author} {\bibfnamefont {J.-C.}\ \bibnamefont {Jeon}}, \bibinfo {author} {\bibfnamefont {A.~K.}\ \bibnamefont {Pandeya}}, \bibinfo {author} {\bibfnamefont {A.}~\bibnamefont {Chakraborty}}, \bibinfo {author} {\bibfnamefont {O.}~\bibnamefont {Busch}}, \bibinfo {author} {\bibfnamefont {A.~K.}\ \bibnamefont {Srivastava}}, \bibinfo {author} {\bibfnamefont {H.}~\bibnamefont {Deniz}}, \bibinfo {author} {\bibfnamefont {J.~M.}\ \bibnamefont {Taylor}}, \emph {et~al.},\ }\href@noop {} {\bibfield  {journal} {\bibinfo  {journal} {Science Advances}\ }\textbf {\bibinfo {volume} {8}},\ \bibinfo {pages} {eabo5930} (\bibinfo {year} {2022})}\BibitemShut {NoStop}%
\bibitem [{\citenamefont {Cao}\ \emph {et~al.}(2023)\citenamefont {Cao}, \citenamefont {Chen}, \citenamefont {Xiao}, \citenamefont {Zhu}, \citenamefont {Yu}, \citenamefont {Wang}, \citenamefont {Qiu}, \citenamefont {Liu}, \citenamefont {Zhao}, \citenamefont {Shao} \emph {et~al.}}]{cao2023anomalous}%
  \BibitemOpen
  \bibfield  {author} {\bibinfo {author} {\bibfnamefont {C.}~\bibnamefont {Cao}}, \bibinfo {author} {\bibfnamefont {S.}~\bibnamefont {Chen}}, \bibinfo {author} {\bibfnamefont {R.-C.}\ \bibnamefont {Xiao}}, \bibinfo {author} {\bibfnamefont {Z.}~\bibnamefont {Zhu}}, \bibinfo {author} {\bibfnamefont {G.}~\bibnamefont {Yu}}, \bibinfo {author} {\bibfnamefont {Y.}~\bibnamefont {Wang}}, \bibinfo {author} {\bibfnamefont {X.}~\bibnamefont {Qiu}}, \bibinfo {author} {\bibfnamefont {L.}~\bibnamefont {Liu}}, \bibinfo {author} {\bibfnamefont {T.}~\bibnamefont {Zhao}}, \bibinfo {author} {\bibfnamefont {D.-F.}\ \bibnamefont {Shao}}, \emph {et~al.},\ }\href@noop {} {\bibfield  {journal} {\bibinfo  {journal} {Nature Communications}\ }\textbf {\bibinfo {volume} {14}},\ \bibinfo {pages} {5873} (\bibinfo {year} {2023})}\BibitemShut {NoStop}%
\bibitem [{\citenamefont {Bai}\ \emph {et~al.}(2021)\citenamefont {Bai}, \citenamefont {Zhou}, \citenamefont {Zhang}, \citenamefont {Kong}, \citenamefont {Liao}, \citenamefont {Feng}, \citenamefont {Chen}, \citenamefont {You}, \citenamefont {Zhou}, \citenamefont {Han},\ and\ \citenamefont {Zhu}}]{Bai2021}%
  \BibitemOpen
  \bibfield  {author} {\bibinfo {author} {\bibfnamefont {H.}~\bibnamefont {Bai}}, \bibinfo {author} {\bibfnamefont {X.~F.}\ \bibnamefont {Zhou}}, \bibinfo {author} {\bibfnamefont {H.~W.}\ \bibnamefont {Zhang}}, \bibinfo {author} {\bibfnamefont {W.~W.}\ \bibnamefont {Kong}}, \bibinfo {author} {\bibfnamefont {L.~Y.}\ \bibnamefont {Liao}}, \bibinfo {author} {\bibfnamefont {X.~Y.}\ \bibnamefont {Feng}}, \bibinfo {author} {\bibfnamefont {X.~Z.}\ \bibnamefont {Chen}}, \bibinfo {author} {\bibfnamefont {Y.~F.}\ \bibnamefont {You}}, \bibinfo {author} {\bibfnamefont {Y.~J.}\ \bibnamefont {Zhou}}, \bibinfo {author} {\bibfnamefont {L.}~\bibnamefont {Han}},\ and\ \bibinfo {author} {\bibfnamefont {W.~X.}\ \bibnamefont {Zhu}},\ }\href@noop {} {\bibfield  {journal} {\bibinfo  {journal} {Physical Review B}\ }\textbf {\bibinfo {volume} {104}},\ \bibinfo {pages} {104401} (\bibinfo {year} {2021})}\BibitemShut {NoStop}%
\bibitem [{\citenamefont {Holanda}\ \emph {et~al.}(2020)\citenamefont {Holanda}, \citenamefont {Saglam}, \citenamefont {Karakas}, \citenamefont {Zang}, \citenamefont {Li}, \citenamefont {Divan}, \citenamefont {Liu}, \citenamefont {Ozatay}, \citenamefont {Novosad}, \citenamefont {Pearson},\ and\ \citenamefont {Hoffmann}}]{Holanda2020}%
  \BibitemOpen
  \bibfield  {author} {\bibinfo {author} {\bibfnamefont {J.}~\bibnamefont {Holanda}}, \bibinfo {author} {\bibfnamefont {H.}~\bibnamefont {Saglam}}, \bibinfo {author} {\bibfnamefont {V.}~\bibnamefont {Karakas}}, \bibinfo {author} {\bibfnamefont {Z.}~\bibnamefont {Zang}}, \bibinfo {author} {\bibfnamefont {Y.}~\bibnamefont {Li}}, \bibinfo {author} {\bibfnamefont {R.}~\bibnamefont {Divan}}, \bibinfo {author} {\bibfnamefont {Y.}~\bibnamefont {Liu}}, \bibinfo {author} {\bibfnamefont {O.}~\bibnamefont {Ozatay}}, \bibinfo {author} {\bibfnamefont {V.}~\bibnamefont {Novosad}}, \bibinfo {author} {\bibfnamefont {J.~E.}\ \bibnamefont {Pearson}},\ and\ \bibinfo {author} {\bibfnamefont {A.}~\bibnamefont {Hoffmann}},\ }\href@noop {} {\bibfield  {journal} {\bibinfo  {journal} {Physical Review Letters}\ }\textbf {\bibinfo {volume} {124}},\ \bibinfo {pages} {087204} (\bibinfo {year} {2020})}\BibitemShut {NoStop}%
\bibitem [{\citenamefont {Zhang}\ \emph {et~al.}(2016)\citenamefont {Zhang}, \citenamefont {Han}, \citenamefont {Yang}, \citenamefont {Sun}, \citenamefont {Zhang}, \citenamefont {Yan},\ and\ \citenamefont {Parkin}}]{zhang2016giant}%
  \BibitemOpen
  \bibfield  {author} {\bibinfo {author} {\bibfnamefont {W.}~\bibnamefont {Zhang}}, \bibinfo {author} {\bibfnamefont {W.}~\bibnamefont {Han}}, \bibinfo {author} {\bibfnamefont {S.-H.}\ \bibnamefont {Yang}}, \bibinfo {author} {\bibfnamefont {Y.}~\bibnamefont {Sun}}, \bibinfo {author} {\bibfnamefont {Y.}~\bibnamefont {Zhang}}, \bibinfo {author} {\bibfnamefont {B.}~\bibnamefont {Yan}},\ and\ \bibinfo {author} {\bibfnamefont {S.~S.}\ \bibnamefont {Parkin}},\ }\href@noop {} {\bibfield  {journal} {\bibinfo  {journal} {Science Advances}\ }\textbf {\bibinfo {volume} {2}},\ \bibinfo {pages} {e1600759} (\bibinfo {year} {2016})}\BibitemShut {NoStop}%
\bibitem [{\citenamefont {Panda}\ \emph {et~al.}(2025)\citenamefont {Panda}, \citenamefont {Mao}, \citenamefont {Peshcherenko}, \citenamefont {Feng}, \citenamefont {Zhang}, \citenamefont {Markou}, \citenamefont {Felser},\ and\ \citenamefont {Lesne}}]{panda2025efficient}%
  \BibitemOpen
  \bibfield  {author} {\bibinfo {author} {\bibfnamefont {S.~N.}\ \bibnamefont {Panda}}, \bibinfo {author} {\bibfnamefont {N.}~\bibnamefont {Mao}}, \bibinfo {author} {\bibfnamefont {N.}~\bibnamefont {Peshcherenko}}, \bibinfo {author} {\bibfnamefont {X.}~\bibnamefont {Feng}}, \bibinfo {author} {\bibfnamefont {Y.}~\bibnamefont {Zhang}}, \bibinfo {author} {\bibfnamefont {A.}~\bibnamefont {Markou}}, \bibinfo {author} {\bibfnamefont {C.}~\bibnamefont {Felser}},\ and\ \bibinfo {author} {\bibfnamefont {E.}~\bibnamefont {Lesne}},\ }\href@noop {} {\bibfield  {journal} {\bibinfo  {journal} {arXiv preprint arXiv:2508.02415}\ } (\bibinfo {year} {2025})}\BibitemShut {NoStop}%
\bibitem [{\citenamefont {Xie}\ \emph {et~al.}(2022)\citenamefont {Xie}, \citenamefont {Chen}, \citenamefont {Zhang}, \citenamefont {Mu}, \citenamefont {Zhang}, \citenamefont {Yan},\ and\ \citenamefont {Wu}}]{xie2022magnetization}%
  \BibitemOpen
  \bibfield  {author} {\bibinfo {author} {\bibfnamefont {H.}~\bibnamefont {Xie}}, \bibinfo {author} {\bibfnamefont {X.}~\bibnamefont {Chen}}, \bibinfo {author} {\bibfnamefont {Q.}~\bibnamefont {Zhang}}, \bibinfo {author} {\bibfnamefont {Z.}~\bibnamefont {Mu}}, \bibinfo {author} {\bibfnamefont {X.}~\bibnamefont {Zhang}}, \bibinfo {author} {\bibfnamefont {B.}~\bibnamefont {Yan}},\ and\ \bibinfo {author} {\bibfnamefont {Y.}~\bibnamefont {Wu}},\ }\href@noop {} {\bibfield  {journal} {\bibinfo  {journal} {Nature Communications}\ }\textbf {\bibinfo {volume} {13}},\ \bibinfo {pages} {5744} (\bibinfo {year} {2022})}\BibitemShut {NoStop}%
\bibitem [{\citenamefont {Yoon}\ \emph {et~al.}(2025)\citenamefont {Yoon}, \citenamefont {Takeuchi}, \citenamefont {Takechi}, \citenamefont {Han}, \citenamefont {Uchimura}, \citenamefont {Yamane}, \citenamefont {Kanai}, \citenamefont {Ieda}, \citenamefont {Ohno},\ and\ \citenamefont {Fukami}}]{yoon2025electrical}%
  \BibitemOpen
  \bibfield  {author} {\bibinfo {author} {\bibfnamefont {J.-Y.}\ \bibnamefont {Yoon}}, \bibinfo {author} {\bibfnamefont {Y.}~\bibnamefont {Takeuchi}}, \bibinfo {author} {\bibfnamefont {R.}~\bibnamefont {Takechi}}, \bibinfo {author} {\bibfnamefont {J.}~\bibnamefont {Han}}, \bibinfo {author} {\bibfnamefont {T.}~\bibnamefont {Uchimura}}, \bibinfo {author} {\bibfnamefont {Y.}~\bibnamefont {Yamane}}, \bibinfo {author} {\bibfnamefont {S.}~\bibnamefont {Kanai}}, \bibinfo {author} {\bibfnamefont {J.}~\bibnamefont {Ieda}}, \bibinfo {author} {\bibfnamefont {H.}~\bibnamefont {Ohno}},\ and\ \bibinfo {author} {\bibfnamefont {S.}~\bibnamefont {Fukami}},\ }\href@noop {} {\bibfield  {journal} {\bibinfo  {journal} {Nature Communications}\ }\textbf {\bibinfo {volume} {16}},\ \bibinfo {pages} {1171} (\bibinfo {year} {2025})}\BibitemShut {NoStop}%
\bibitem [{\citenamefont {Rimmler}\ \emph {et~al.}(2025)\citenamefont {Rimmler}, \citenamefont {Pal},\ and\ \citenamefont {Parkin}}]{rimmler2025non}%
  \BibitemOpen
  \bibfield  {author} {\bibinfo {author} {\bibfnamefont {B.~H.}\ \bibnamefont {Rimmler}}, \bibinfo {author} {\bibfnamefont {B.}~\bibnamefont {Pal}},\ and\ \bibinfo {author} {\bibfnamefont {S.~S.}\ \bibnamefont {Parkin}},\ }\href@noop {} {\bibfield  {journal} {\bibinfo  {journal} {Nature Reviews Materials}\ }\textbf {\bibinfo {volume} {10}},\ \bibinfo {pages} {109} (\bibinfo {year} {2025})}\BibitemShut {NoStop}%
\bibitem [{\citenamefont {Gurung}\ \emph {et~al.}(2019)\citenamefont {Gurung}, \citenamefont {Shao}, \citenamefont {Paudel},\ and\ \citenamefont {Tsymbal}}]{gurung2019anomalous}%
  \BibitemOpen
  \bibfield  {author} {\bibinfo {author} {\bibfnamefont {G.}~\bibnamefont {Gurung}}, \bibinfo {author} {\bibfnamefont {D.-F.}\ \bibnamefont {Shao}}, \bibinfo {author} {\bibfnamefont {T.~R.}\ \bibnamefont {Paudel}},\ and\ \bibinfo {author} {\bibfnamefont {E.~Y.}\ \bibnamefont {Tsymbal}},\ }\href@noop {} {\bibfield  {journal} {\bibinfo  {journal} {Physical Review Materials}\ }\textbf {\bibinfo {volume} {3}},\ \bibinfo {pages} {044409} (\bibinfo {year} {2019})}\BibitemShut {NoStop}%
\bibitem [{\citenamefont {Kota}\ \emph {et~al.}(2008)\citenamefont {Kota}, \citenamefont {Tsuchiura},\ and\ \citenamefont {Sakuma}}]{kota2008ab}%
  \BibitemOpen
  \bibfield  {author} {\bibinfo {author} {\bibfnamefont {Y.}~\bibnamefont {Kota}}, \bibinfo {author} {\bibfnamefont {H.}~\bibnamefont {Tsuchiura}},\ and\ \bibinfo {author} {\bibfnamefont {A.}~\bibnamefont {Sakuma}},\ }\href@noop {} {\bibfield  {journal} {\bibinfo  {journal} {IEEE Transactions on Magnetics}\ }\textbf {\bibinfo {volume} {44}},\ \bibinfo {pages} {3131} (\bibinfo {year} {2008})}\BibitemShut {NoStop}%
\bibitem [{\citenamefont {Sinha}\ \emph {et~al.}(2025{\natexlab{a}})\citenamefont {Sinha}, \citenamefont {Sachin}, \citenamefont {Sinha}, \citenamefont {Roy}, \citenamefont {Kanungo},\ and\ \citenamefont {Manna}}]{sinha2025occurrence}%
  \BibitemOpen
  \bibfield  {author} {\bibinfo {author} {\bibfnamefont {I.}~\bibnamefont {Sinha}}, \bibinfo {author} {\bibfnamefont {S.}~\bibnamefont {Sachin}}, \bibinfo {author} {\bibfnamefont {S.}~\bibnamefont {Sinha}}, \bibinfo {author} {\bibfnamefont {R.}~\bibnamefont {Roy}}, \bibinfo {author} {\bibfnamefont {S.}~\bibnamefont {Kanungo}},\ and\ \bibinfo {author} {\bibfnamefont {S.}~\bibnamefont {Manna}},\ }\href@noop {} {\bibfield  {journal} {\bibinfo  {journal} {Physical Review Materials}\ }\textbf {\bibinfo {volume} {9}},\ \bibinfo {pages} {074202} (\bibinfo {year} {2025}{\natexlab{a}})}\BibitemShut {NoStop}%
\bibitem [{\citenamefont {Sinha}\ \emph {et~al.}(2025{\natexlab{b}})\citenamefont {Sinha}, \citenamefont {Sinha}, \citenamefont {Naskar},\ and\ \citenamefont {Manna}}]{sinha2025magnetic}%
  \BibitemOpen
  \bibfield  {author} {\bibinfo {author} {\bibfnamefont {I.}~\bibnamefont {Sinha}}, \bibinfo {author} {\bibfnamefont {S.}~\bibnamefont {Sinha}}, \bibinfo {author} {\bibfnamefont {S.}~\bibnamefont {Naskar}},\ and\ \bibinfo {author} {\bibfnamefont {S.}~\bibnamefont {Manna}},\ }\href@noop {} {\bibfield  {journal} {\bibinfo  {journal} {Journal of Physics: Condensed Matter}\ }\textbf {\bibinfo {volume} {37}},\ \bibinfo {pages} {115001} (\bibinfo {year} {2025}{\natexlab{b}})}\BibitemShut {NoStop}%
\bibitem [{\citenamefont {Sinha}\ \emph {et~al.}(2025{\natexlab{c}})\citenamefont {Sinha}, \citenamefont {Dutta}, \citenamefont {Firdosh}, \citenamefont {Sinha}, \citenamefont {Ganguli},\ and\ \citenamefont {Manna}}]{sinha2025anomalous}%
  \BibitemOpen
  \bibfield  {author} {\bibinfo {author} {\bibfnamefont {I.}~\bibnamefont {Sinha}}, \bibinfo {author} {\bibfnamefont {P.}~\bibnamefont {Dutta}}, \bibinfo {author} {\bibfnamefont {N.}~\bibnamefont {Firdosh}}, \bibinfo {author} {\bibfnamefont {S.}~\bibnamefont {Sinha}}, \bibinfo {author} {\bibfnamefont {N.}~\bibnamefont {Ganguli}},\ and\ \bibinfo {author} {\bibfnamefont {S.}~\bibnamefont {Manna}},\ }\href@noop {} {\bibfield  {journal} {\bibinfo  {journal} {Journal of Physics: Condensed Matter}\ } (\bibinfo {year} {2025}{\natexlab{c}})}\BibitemShut {NoStop}%
\bibitem [{\citenamefont {Liu}\ \emph {et~al.}(2018)\citenamefont {Liu}, \citenamefont {Chen}, \citenamefont {Wang}, \citenamefont {Liu}, \citenamefont {Wang}, \citenamefont {Feng}, \citenamefont {Yan}, \citenamefont {Wang}, \citenamefont {Jiang}, \citenamefont {Coey} \emph {et~al.}}]{liu2018electrical}%
  \BibitemOpen
  \bibfield  {author} {\bibinfo {author} {\bibfnamefont {Z.}~\bibnamefont {Liu}}, \bibinfo {author} {\bibfnamefont {H.}~\bibnamefont {Chen}}, \bibinfo {author} {\bibfnamefont {J.}~\bibnamefont {Wang}}, \bibinfo {author} {\bibfnamefont {J.}~\bibnamefont {Liu}}, \bibinfo {author} {\bibfnamefont {K.}~\bibnamefont {Wang}}, \bibinfo {author} {\bibfnamefont {Z.}~\bibnamefont {Feng}}, \bibinfo {author} {\bibfnamefont {H.}~\bibnamefont {Yan}}, \bibinfo {author} {\bibfnamefont {X.}~\bibnamefont {Wang}}, \bibinfo {author} {\bibfnamefont {C.}~\bibnamefont {Jiang}}, \bibinfo {author} {\bibfnamefont {J.}~\bibnamefont {Coey}}, \emph {et~al.},\ }\href@noop {} {\bibfield  {journal} {\bibinfo  {journal} {Nature Electronics}\ }\textbf {\bibinfo {volume} {1}},\ \bibinfo {pages} {172} (\bibinfo {year} {2018})}\BibitemShut {NoStop}%
\bibitem [{\citenamefont {Kumar}\ \emph {et~al.}(2024)\citenamefont {Kumar}, \citenamefont {Kumar}, \citenamefont {Sharma}, \citenamefont {Khanna},\ and\ \citenamefont {Kuanr}}]{kumar2024influence}%
  \BibitemOpen
  \bibfield  {author} {\bibinfo {author} {\bibfnamefont {P.}~\bibnamefont {Kumar}}, \bibinfo {author} {\bibfnamefont {R.}~\bibnamefont {Kumar}}, \bibinfo {author} {\bibfnamefont {V.}~\bibnamefont {Sharma}}, \bibinfo {author} {\bibfnamefont {M.~K.}\ \bibnamefont {Khanna}},\ and\ \bibinfo {author} {\bibfnamefont {B.~K.}\ \bibnamefont {Kuanr}},\ }\href@noop {} {\bibfield  {journal} {\bibinfo  {journal} {Journal of Alloys and Compounds}\ }\textbf {\bibinfo {volume} {988}},\ \bibinfo {pages} {174314} (\bibinfo {year} {2024})}\BibitemShut {NoStop}%
\bibitem [{\citenamefont {Kharmouche}\ and\ \citenamefont {Cherrad}(2024)}]{kharmouche2024structural}%
  \BibitemOpen
  \bibfield  {author} {\bibinfo {author} {\bibfnamefont {A.}~\bibnamefont {Kharmouche}}\ and\ \bibinfo {author} {\bibfnamefont {O.}~\bibnamefont {Cherrad}},\ }\href@noop {} {\bibfield  {journal} {\bibinfo  {journal} {Physica B: Condensed Matter}\ }\textbf {\bibinfo {volume} {680}},\ \bibinfo {pages} {415803} (\bibinfo {year} {2024})}\BibitemShut {NoStop}%
\bibitem [{\citenamefont {Chen}\ \emph {et~al.}(2014)\citenamefont {Chen}, \citenamefont {Niu},\ and\ \citenamefont {MacDonald}}]{chen2014anomalous}%
  \BibitemOpen
  \bibfield  {author} {\bibinfo {author} {\bibfnamefont {H.}~\bibnamefont {Chen}}, \bibinfo {author} {\bibfnamefont {Q.}~\bibnamefont {Niu}},\ and\ \bibinfo {author} {\bibfnamefont {A.~H.}\ \bibnamefont {MacDonald}},\ }\href@noop {} {\bibfield  {journal} {\bibinfo  {journal} {Physical Review Letters}\ }\textbf {\bibinfo {volume} {112}},\ \bibinfo {pages} {017205} (\bibinfo {year} {2014})}\BibitemShut {NoStop}%
\bibitem [{\citenamefont {Xu}\ \emph {et~al.}(2024)\citenamefont {Xu}, \citenamefont {Dai}, \citenamefont {Jiang}, \citenamefont {Xiong}, \citenamefont {Cheng}, \citenamefont {Tai}, \citenamefont {Tang}, \citenamefont {Sun}, \citenamefont {He}, \citenamefont {Yang} \emph {et~al.}}]{xu2024universal}%
  \BibitemOpen
  \bibfield  {author} {\bibinfo {author} {\bibfnamefont {S.}~\bibnamefont {Xu}}, \bibinfo {author} {\bibfnamefont {B.}~\bibnamefont {Dai}}, \bibinfo {author} {\bibfnamefont {Y.}~\bibnamefont {Jiang}}, \bibinfo {author} {\bibfnamefont {D.}~\bibnamefont {Xiong}}, \bibinfo {author} {\bibfnamefont {H.}~\bibnamefont {Cheng}}, \bibinfo {author} {\bibfnamefont {L.}~\bibnamefont {Tai}}, \bibinfo {author} {\bibfnamefont {M.}~\bibnamefont {Tang}}, \bibinfo {author} {\bibfnamefont {Y.}~\bibnamefont {Sun}}, \bibinfo {author} {\bibfnamefont {Y.}~\bibnamefont {He}}, \bibinfo {author} {\bibfnamefont {B.}~\bibnamefont {Yang}}, \emph {et~al.},\ }\href@noop {} {\bibfield  {journal} {\bibinfo  {journal} {Nature Communications}\ }\textbf {\bibinfo {volume} {15}},\ \bibinfo {pages} {3717} (\bibinfo {year} {2024})}\BibitemShut {NoStop}%
\bibitem [{zzS()}]{zzSI}%
  \BibitemOpen
  \href@noop {} {}\bibinfo {note} {See Supplemental Material for temperature-dependent longitudinal resistivity, magnetoresistance, transverse resistivity.}\BibitemShut {Stop}%
\bibitem [{\citenamefont {Dong}\ \emph {et~al.}(2021)\citenamefont {Dong}, \citenamefont {Jiao}, \citenamefont {Yuan}, \citenamefont {Sun}, \citenamefont {He}, \citenamefont {Jin}, \citenamefont {Mo},\ and\ \citenamefont {Song}}]{dong2021low}%
  \BibitemOpen
  \bibfield  {author} {\bibinfo {author} {\bibfnamefont {K.}~\bibnamefont {Dong}}, \bibinfo {author} {\bibfnamefont {Y.}~\bibnamefont {Jiao}}, \bibinfo {author} {\bibfnamefont {Z.}~\bibnamefont {Yuan}}, \bibinfo {author} {\bibfnamefont {C.}~\bibnamefont {Sun}}, \bibinfo {author} {\bibfnamefont {K.}~\bibnamefont {He}}, \bibinfo {author} {\bibfnamefont {F.}~\bibnamefont {Jin}}, \bibinfo {author} {\bibfnamefont {W.}~\bibnamefont {Mo}},\ and\ \bibinfo {author} {\bibfnamefont {J.}~\bibnamefont {Song}},\ }\href@noop {} {\bibfield  {journal} {\bibinfo  {journal} {Journal of Magnetism and Magnetic Materials}\ }\textbf {\bibinfo {volume} {523}},\ \bibinfo {pages} {167615} (\bibinfo {year} {2021})}\BibitemShut {NoStop}%
\bibitem [{\citenamefont {Dürrenfeld}\ \emph {et~al.}(2015)\citenamefont {Dürrenfeld}, \citenamefont {Gerhard}, \citenamefont {Chico}, \citenamefont {Dumas}, \citenamefont {Ranjbar}, \citenamefont {Bergman}, \citenamefont {Bergqvist}, \citenamefont {Delin}, \citenamefont {Gould}, \citenamefont {Molenkamp},\ and\ \citenamefont {Åkerman}}]{Durrenfeld2015}%
  \BibitemOpen
  \bibfield  {author} {\bibinfo {author} {\bibfnamefont {P.}~\bibnamefont {Dürrenfeld}}, \bibinfo {author} {\bibfnamefont {F.}~\bibnamefont {Gerhard}}, \bibinfo {author} {\bibfnamefont {J.}~\bibnamefont {Chico}}, \bibinfo {author} {\bibfnamefont {R.}~\bibnamefont {Dumas}}, \bibinfo {author} {\bibfnamefont {M.}~\bibnamefont {Ranjbar}}, \bibinfo {author} {\bibfnamefont {A.}~\bibnamefont {Bergman}}, \bibinfo {author} {\bibfnamefont {L.}~\bibnamefont {Bergqvist}}, \bibinfo {author} {\bibfnamefont {A.}~\bibnamefont {Delin}}, \bibinfo {author} {\bibfnamefont {C.}~\bibnamefont {Gould}}, \bibinfo {author} {\bibfnamefont {L.}~\bibnamefont {Molenkamp}},\ and\ \bibinfo {author} {\bibfnamefont {J.}~\bibnamefont {Åkerman}},\ }\href@noop {} {\bibfield  {journal} {\bibinfo  {journal} {Physical Review B}\ }\textbf {\bibinfo {volume} {92}},\ \bibinfo {pages} {214424} (\bibinfo {year} {2015})}\BibitemShut {NoStop}%
\bibitem [{\citenamefont {Jungfleisch}\ \emph {et~al.}(2015)\citenamefont {Jungfleisch}, \citenamefont {Chumak}, \citenamefont {Kehlberger}, \citenamefont {Lauer}, \citenamefont {Kim}, \citenamefont {Onbasli}, \citenamefont {Ross}, \citenamefont {Kläui},\ and\ \citenamefont {Hillebrands}}]{Jungfleisch2015}%
  \BibitemOpen
  \bibfield  {author} {\bibinfo {author} {\bibfnamefont {M.~B.}\ \bibnamefont {Jungfleisch}}, \bibinfo {author} {\bibfnamefont {A.~V.}\ \bibnamefont {Chumak}}, \bibinfo {author} {\bibfnamefont {A.}~\bibnamefont {Kehlberger}}, \bibinfo {author} {\bibfnamefont {V.}~\bibnamefont {Lauer}}, \bibinfo {author} {\bibfnamefont {D.~H.}\ \bibnamefont {Kim}}, \bibinfo {author} {\bibfnamefont {M.~C.}\ \bibnamefont {Onbasli}}, \bibinfo {author} {\bibfnamefont {C.~A.}\ \bibnamefont {Ross}}, \bibinfo {author} {\bibfnamefont {M.}~\bibnamefont {Kläui}},\ and\ \bibinfo {author} {\bibfnamefont {B.}~\bibnamefont {Hillebrands}},\ }\href@noop {} {\bibfield  {journal} {\bibinfo  {journal} {Physical Review B}\ }\textbf {\bibinfo {volume} {91}},\ \bibinfo {pages} {134407} (\bibinfo {year} {2015})}\BibitemShut {NoStop}%
\bibitem [{\citenamefont {Sun}\ \emph {et~al.}(2020)\citenamefont {Sun}, \citenamefont {Li}, \citenamefont {Xie}, \citenamefont {Li}, \citenamefont {Zhao}, \citenamefont {Liu}, \citenamefont {Zhang}, \citenamefont {Zhu}, \citenamefont {Cheng},\ and\ \citenamefont {He}}]{Sun2020}%
  \BibitemOpen
  \bibfield  {author} {\bibinfo {author} {\bibfnamefont {R.}~\bibnamefont {Sun}}, \bibinfo {author} {\bibfnamefont {Y.}~\bibnamefont {Li}}, \bibinfo {author} {\bibfnamefont {Z.~K.}\ \bibnamefont {Xie}}, \bibinfo {author} {\bibfnamefont {Y.}~\bibnamefont {Li}}, \bibinfo {author} {\bibfnamefont {X.-T.}\ \bibnamefont {Zhao}}, \bibinfo {author} {\bibfnamefont {W.}~\bibnamefont {Liu}}, \bibinfo {author} {\bibfnamefont {Z.~D.}\ \bibnamefont {Zhang}}, \bibinfo {author} {\bibfnamefont {T.}~\bibnamefont {Zhu}}, \bibinfo {author} {\bibfnamefont {Z.-H.}\ \bibnamefont {Cheng}},\ and\ \bibinfo {author} {\bibfnamefont {W.}~\bibnamefont {He}},\ }\href@noop {} {\bibfield  {journal} {\bibinfo  {journal} {Journal of Magnetism and Magnetic Materials}\ }\textbf {\bibinfo {volume} {497}},\ \bibinfo {pages} {165971} (\bibinfo {year} {2020})}\BibitemShut {NoStop}%
\bibitem [{\citenamefont {Nibarger}\ \emph {et~al.}(2003)\citenamefont {Nibarger}, \citenamefont {Lopusnik},\ and\ \citenamefont {Silva}}]{Nibarger2003}%
  \BibitemOpen
  \bibfield  {author} {\bibinfo {author} {\bibfnamefont {J.~P.}\ \bibnamefont {Nibarger}}, \bibinfo {author} {\bibfnamefont {R.}~\bibnamefont {Lopusnik}},\ and\ \bibinfo {author} {\bibfnamefont {T.~J.}\ \bibnamefont {Silva}},\ }\href@noop {} {\bibfield  {journal} {\bibinfo  {journal} {Applied Physics Letters}\ }\textbf {\bibinfo {volume} {82}},\ \bibinfo {pages} {2112} (\bibinfo {year} {2003})}\BibitemShut {NoStop}%
\bibitem [{\citenamefont {Hazra}\ \emph {et~al.}(2019)\citenamefont {Hazra}, \citenamefont {Kaul}, \citenamefont {Srinath},\ and\ \citenamefont {Raja}}]{Hazra2019}%
  \BibitemOpen
  \bibfield  {author} {\bibinfo {author} {\bibfnamefont {B.~K.}\ \bibnamefont {Hazra}}, \bibinfo {author} {\bibfnamefont {S.~N.}\ \bibnamefont {Kaul}}, \bibinfo {author} {\bibfnamefont {S.}~\bibnamefont {Srinath}},\ and\ \bibinfo {author} {\bibfnamefont {M.~M.}\ \bibnamefont {Raja}},\ }\href@noop {} {\bibfield  {journal} {\bibinfo  {journal} {Journal of Physics D: Applied Physics}\ }\textbf {\bibinfo {volume} {52}},\ \bibinfo {pages} {325002} (\bibinfo {year} {2019})}\BibitemShut {NoStop}%
\bibitem [{\citenamefont {Inaba}\ \emph {et~al.}(2006)\citenamefont {Inaba}, \citenamefont {Asanuma}, \citenamefont {Igarashi}, \citenamefont {Mori}, \citenamefont {Kirino}, \citenamefont {Koike},\ and\ \citenamefont {Morita}}]{Inaba2006}%
  \BibitemOpen
  \bibfield  {author} {\bibinfo {author} {\bibfnamefont {N.}~\bibnamefont {Inaba}}, \bibinfo {author} {\bibfnamefont {H.}~\bibnamefont {Asanuma}}, \bibinfo {author} {\bibfnamefont {S.}~\bibnamefont {Igarashi}}, \bibinfo {author} {\bibfnamefont {S.}~\bibnamefont {Mori}}, \bibinfo {author} {\bibfnamefont {F.}~\bibnamefont {Kirino}}, \bibinfo {author} {\bibfnamefont {K.}~\bibnamefont {Koike}},\ and\ \bibinfo {author} {\bibfnamefont {H.}~\bibnamefont {Morita}},\ }\href@noop {} {\bibfield  {journal} {\bibinfo  {journal} {IEEE Transactions on Magnetics}\ }\textbf {\bibinfo {volume} {42}},\ \bibinfo {pages} {2372} (\bibinfo {year} {2006})}\BibitemShut {NoStop}%
\bibitem [{\citenamefont {Schulz}\ \emph {et~al.}(2021)\citenamefont {Schulz}, \citenamefont {Lawitzki}, \citenamefont {G{\l}owi{\'n}ski}, \citenamefont {Lisiecki}, \citenamefont {Tr{\"a}ger}, \citenamefont {Ku{\'s}wik}, \citenamefont {Goering}, \citenamefont {Sch{\"u}tz},\ and\ \citenamefont {Gr{\"a}fe}}]{schulz2021increase}%
  \BibitemOpen
  \bibfield  {author} {\bibinfo {author} {\bibfnamefont {F.}~\bibnamefont {Schulz}}, \bibinfo {author} {\bibfnamefont {R.}~\bibnamefont {Lawitzki}}, \bibinfo {author} {\bibfnamefont {H.}~\bibnamefont {G{\l}owi{\'n}ski}}, \bibinfo {author} {\bibfnamefont {F.}~\bibnamefont {Lisiecki}}, \bibinfo {author} {\bibfnamefont {N.}~\bibnamefont {Tr{\"a}ger}}, \bibinfo {author} {\bibfnamefont {P.}~\bibnamefont {Ku{\'s}wik}}, \bibinfo {author} {\bibfnamefont {E.}~\bibnamefont {Goering}}, \bibinfo {author} {\bibfnamefont {G.}~\bibnamefont {Sch{\"u}tz}},\ and\ \bibinfo {author} {\bibfnamefont {J.}~\bibnamefont {Gr{\"a}fe}},\ }\href@noop {} {\bibfield  {journal} {\bibinfo  {journal} {Journal of Applied Physics}\ }\textbf {\bibinfo {volume} {129}} (\bibinfo {year} {2021})}\BibitemShut {NoStop}%
\bibitem [{\citenamefont {Manschot}\ \emph {et~al.}(2004)\citenamefont {Manschot}, \citenamefont {Brataas},\ and\ \citenamefont {Bauer}}]{Manschot2004}%
  \BibitemOpen
  \bibfield  {author} {\bibinfo {author} {\bibfnamefont {J.}~\bibnamefont {Manschot}}, \bibinfo {author} {\bibfnamefont {A.}~\bibnamefont {Brataas}},\ and\ \bibinfo {author} {\bibfnamefont {G.~E.~W.}\ \bibnamefont {Bauer}},\ }\href@noop {} {\bibfield  {journal} {\bibinfo  {journal} {Applied Physics Letters}\ }\textbf {\bibinfo {volume} {85}},\ \bibinfo {pages} {3250} (\bibinfo {year} {2004})}\BibitemShut {NoStop}%
\bibitem [{\citenamefont {Lund}\ \emph {et~al.}(2021)\citenamefont {Lund}, \citenamefont {Salimath},\ and\ \citenamefont {Hals}}]{lund2021spin}%
  \BibitemOpen
  \bibfield  {author} {\bibinfo {author} {\bibfnamefont {M.~A.}\ \bibnamefont {Lund}}, \bibinfo {author} {\bibfnamefont {A.}~\bibnamefont {Salimath}},\ and\ \bibinfo {author} {\bibfnamefont {K.~M.}\ \bibnamefont {Hals}},\ }\href@noop {} {\bibfield  {journal} {\bibinfo  {journal} {Physical Review B}\ }\textbf {\bibinfo {volume} {104}},\ \bibinfo {pages} {174424} (\bibinfo {year} {2021})}\BibitemShut {NoStop}%
\bibitem [{\citenamefont {Frangou}\ \emph {et~al.}(2016)\citenamefont {Frangou}, \citenamefont {Oyarzun}, \citenamefont {Auffret}, \citenamefont {Vila}, \citenamefont {Gambarelli},\ and\ \citenamefont {Baltz}}]{frangou2016enhanced}%
  \BibitemOpen
  \bibfield  {author} {\bibinfo {author} {\bibfnamefont {L.}~\bibnamefont {Frangou}}, \bibinfo {author} {\bibfnamefont {S.}~\bibnamefont {Oyarzun}}, \bibinfo {author} {\bibfnamefont {S.}~\bibnamefont {Auffret}}, \bibinfo {author} {\bibfnamefont {L.}~\bibnamefont {Vila}}, \bibinfo {author} {\bibfnamefont {S.}~\bibnamefont {Gambarelli}},\ and\ \bibinfo {author} {\bibfnamefont {V.}~\bibnamefont {Baltz}},\ }\href@noop {} {\bibfield  {journal} {\bibinfo  {journal} {Physical Review Letters}\ }\textbf {\bibinfo {volume} {116}},\ \bibinfo {pages} {077203} (\bibinfo {year} {2016})}\BibitemShut {NoStop}%
\bibitem [{\citenamefont {Pal}\ \emph {et~al.}(2024)\citenamefont {Pal}, \citenamefont {Nandi}, \citenamefont {Nath}, \citenamefont {Pal}, \citenamefont {Sharma}, \citenamefont {Manna}, \citenamefont {Barman},\ and\ \citenamefont {Mitra}}]{Pal2024}%
  \BibitemOpen
  \bibfield  {author} {\bibinfo {author} {\bibfnamefont {S.}~\bibnamefont {Pal}}, \bibinfo {author} {\bibfnamefont {A.}~\bibnamefont {Nandi}}, \bibinfo {author} {\bibfnamefont {S.~G.}\ \bibnamefont {Nath}}, \bibinfo {author} {\bibfnamefont {P.~K.}\ \bibnamefont {Pal}}, \bibinfo {author} {\bibfnamefont {K.}~\bibnamefont {Sharma}}, \bibinfo {author} {\bibfnamefont {S.}~\bibnamefont {Manna}}, \bibinfo {author} {\bibfnamefont {A.}~\bibnamefont {Barman}},\ and\ \bibinfo {author} {\bibfnamefont {C.}~\bibnamefont {Mitra}},\ }\href@noop {} {\bibfield  {journal} {\bibinfo  {journal} {Appl. Phys. Lett.}\ }\textbf {\bibinfo {volume} {124}},\ \bibinfo {pages} {112403} (\bibinfo {year} {2024})}\BibitemShut {NoStop}%
\bibitem [{\citenamefont {Khan}\ \emph {et~al.}(2024)\citenamefont {Khan}, \citenamefont {Kumar}, \citenamefont {Gupta}, \citenamefont {Yadav}, \citenamefont {{\AA}kerman},\ and\ \citenamefont {Muduli}}]{khan2024magnetodynamic}%
  \BibitemOpen
  \bibfield  {author} {\bibinfo {author} {\bibfnamefont {K.~I.~A.}\ \bibnamefont {Khan}}, \bibinfo {author} {\bibfnamefont {A.}~\bibnamefont {Kumar}}, \bibinfo {author} {\bibfnamefont {P.}~\bibnamefont {Gupta}}, \bibinfo {author} {\bibfnamefont {R.~S.}\ \bibnamefont {Yadav}}, \bibinfo {author} {\bibfnamefont {J.}~\bibnamefont {{\AA}kerman}},\ and\ \bibinfo {author} {\bibfnamefont {P.~K.}\ \bibnamefont {Muduli}},\ }\href@noop {} {\bibfield  {journal} {\bibinfo  {journal} {Scientific Reports}\ }\textbf {\bibinfo {volume} {14}},\ \bibinfo {pages} {3487} (\bibinfo {year} {2024})}\BibitemShut {NoStop}%
\bibitem [{\citenamefont {Ding}\ \emph {et~al.}(2024)\citenamefont {Ding}, \citenamefont {Wang}, \citenamefont {Legrand}, \citenamefont {No{\"e}l},\ and\ \citenamefont {Gambardella}}]{ding2024mitigation}%
  \BibitemOpen
  \bibfield  {author} {\bibinfo {author} {\bibfnamefont {S.}~\bibnamefont {Ding}}, \bibinfo {author} {\bibfnamefont {H.}~\bibnamefont {Wang}}, \bibinfo {author} {\bibfnamefont {W.}~\bibnamefont {Legrand}}, \bibinfo {author} {\bibfnamefont {P.}~\bibnamefont {No{\"e}l}},\ and\ \bibinfo {author} {\bibfnamefont {P.}~\bibnamefont {Gambardella}},\ }\href@noop {} {\bibfield  {journal} {\bibinfo  {journal} {Nano Letters}\ }\textbf {\bibinfo {volume} {24}},\ \bibinfo {pages} {10251} (\bibinfo {year} {2024})}\BibitemShut {NoStop}%
\bibitem [{\citenamefont {Mosendz}\ \emph {et~al.}(2010)\citenamefont {Mosendz}, \citenamefont {Pearson}, \citenamefont {Fradin}, \citenamefont {Bauer}, \citenamefont {Bader},\ and\ \citenamefont {Hoffmann}}]{mosendz2010quantifying}%
  \BibitemOpen
  \bibfield  {author} {\bibinfo {author} {\bibfnamefont {O.}~\bibnamefont {Mosendz}}, \bibinfo {author} {\bibfnamefont {J.}~\bibnamefont {Pearson}}, \bibinfo {author} {\bibfnamefont {F.}~\bibnamefont {Fradin}}, \bibinfo {author} {\bibfnamefont {G.}~\bibnamefont {Bauer}}, \bibinfo {author} {\bibfnamefont {S.}~\bibnamefont {Bader}},\ and\ \bibinfo {author} {\bibfnamefont {A.}~\bibnamefont {Hoffmann}},\ }\href@noop {} {\bibfield  {journal} {\bibinfo  {journal} {Physical Review Letters}\ }\textbf {\bibinfo {volume} {104}},\ \bibinfo {pages} {046601} (\bibinfo {year} {2010})}\BibitemShut {NoStop}%
\bibitem [{\citenamefont {Kimata}\ \emph {et~al.}(2019)\citenamefont {Kimata}, \citenamefont {Chen}, \citenamefont {Kondou}, \citenamefont {Sugimoto}, \citenamefont {Muduli}, \citenamefont {Ikhlas}, \citenamefont {Omori}, \citenamefont {Tomita}, \citenamefont {MacDonald}, \citenamefont {Nakatsuji} \emph {et~al.}}]{kimata2019magnetic}%
  \BibitemOpen
  \bibfield  {author} {\bibinfo {author} {\bibfnamefont {M.}~\bibnamefont {Kimata}}, \bibinfo {author} {\bibfnamefont {H.}~\bibnamefont {Chen}}, \bibinfo {author} {\bibfnamefont {K.}~\bibnamefont {Kondou}}, \bibinfo {author} {\bibfnamefont {S.}~\bibnamefont {Sugimoto}}, \bibinfo {author} {\bibfnamefont {P.~K.}\ \bibnamefont {Muduli}}, \bibinfo {author} {\bibfnamefont {M.}~\bibnamefont {Ikhlas}}, \bibinfo {author} {\bibfnamefont {Y.}~\bibnamefont {Omori}}, \bibinfo {author} {\bibfnamefont {T.}~\bibnamefont {Tomita}}, \bibinfo {author} {\bibfnamefont {A.~H.}\ \bibnamefont {MacDonald}}, \bibinfo {author} {\bibfnamefont {S.}~\bibnamefont {Nakatsuji}}, \emph {et~al.},\ }\href@noop {} {\bibfield  {journal} {\bibinfo  {journal} {Nature}\ }\textbf {\bibinfo {volume} {565}},\ \bibinfo {pages} {627} (\bibinfo {year} {2019})}\BibitemShut {NoStop}%
\bibitem [{\citenamefont {Hayashi}\ \emph {et~al.}(2021)\citenamefont {Hayashi}, \citenamefont {Musha}, \citenamefont {Sakimura},\ and\ \citenamefont {Ando}}]{hayashi2021spin}%
  \BibitemOpen
  \bibfield  {author} {\bibinfo {author} {\bibfnamefont {H.}~\bibnamefont {Hayashi}}, \bibinfo {author} {\bibfnamefont {A.}~\bibnamefont {Musha}}, \bibinfo {author} {\bibfnamefont {H.}~\bibnamefont {Sakimura}},\ and\ \bibinfo {author} {\bibfnamefont {K.}~\bibnamefont {Ando}},\ }\href@noop {} {\bibfield  {journal} {\bibinfo  {journal} {Physical Review Research}\ }\textbf {\bibinfo {volume} {3}},\ \bibinfo {pages} {013042} (\bibinfo {year} {2021})}\BibitemShut {NoStop}%
\bibitem [{\citenamefont {Wang}\ \emph {et~al.}(2014)\citenamefont {Wang}, \citenamefont {Du}, \citenamefont {Pu}, \citenamefont {Adur}, \citenamefont {Hammel},\ and\ \citenamefont {Yang}}]{Wang2014}%
  \BibitemOpen
  \bibfield  {author} {\bibinfo {author} {\bibfnamefont {H.~L.}\ \bibnamefont {Wang}}, \bibinfo {author} {\bibfnamefont {C.~H.}\ \bibnamefont {Du}}, \bibinfo {author} {\bibfnamefont {Y.}~\bibnamefont {Pu}}, \bibinfo {author} {\bibfnamefont {R.}~\bibnamefont {Adur}}, \bibinfo {author} {\bibfnamefont {P.~C.}\ \bibnamefont {Hammel}},\ and\ \bibinfo {author} {\bibfnamefont {F.~Y.}\ \bibnamefont {Yang}},\ }\href@noop {} {\bibfield  {journal} {\bibinfo  {journal} {Physical Review Letters}\ }\textbf {\bibinfo {volume} {112}},\ \bibinfo {pages} {197201} (\bibinfo {year} {2014})}\BibitemShut {NoStop}%
\bibitem [{\citenamefont {Kumar}\ \emph {et~al.}(2025)\citenamefont {Kumar}, \citenamefont {Sharma}, \citenamefont {Khanna},\ and\ \citenamefont {Kuanr}}]{kumar2025effect}%
  \BibitemOpen
  \bibfield  {author} {\bibinfo {author} {\bibfnamefont {P.}~\bibnamefont {Kumar}}, \bibinfo {author} {\bibfnamefont {V.}~\bibnamefont {Sharma}}, \bibinfo {author} {\bibfnamefont {M.~K.}\ \bibnamefont {Khanna}},\ and\ \bibinfo {author} {\bibfnamefont {B.~K.}\ \bibnamefont {Kuanr}},\ }\href@noop {} {\bibfield  {journal} {\bibinfo  {journal} {Physics Letters A}\ ,\ \bibinfo {pages} {130722}} (\bibinfo {year} {2025})}\BibitemShut {NoStop}%
\bibitem [{\citenamefont {Sharma}\ \emph {et~al.}(2022)\citenamefont {Sharma}, \citenamefont {Sharma}, \citenamefont {Ghosh},\ and\ \citenamefont {Kuanr}}]{sharma2022magnetization}%
  \BibitemOpen
  \bibfield  {author} {\bibinfo {author} {\bibfnamefont {V.}~\bibnamefont {Sharma}}, \bibinfo {author} {\bibfnamefont {V.}~\bibnamefont {Sharma}}, \bibinfo {author} {\bibfnamefont {R.~K.}\ \bibnamefont {Ghosh}},\ and\ \bibinfo {author} {\bibfnamefont {B.~K.}\ \bibnamefont {Kuanr}},\ }\href@noop {} {\bibfield  {journal} {\bibinfo  {journal} {Journal of Applied Physics}\ }\textbf {\bibinfo {volume} {132}} (\bibinfo {year} {2022})}\BibitemShut {NoStop}%
\end{thebibliography}%

\end{document}